\documentclass[%
 reprint,
 amsmath,amssymb,
 aps,
]{revtex4-2}

\usepackage{amsthm}

\newtheorem{theorem}{Theorem}      % numbered: Theorem 1,2,3,...

\newtheorem{corollary}{Corollary}
\newtheorem{definition}{Definition}

\usepackage{soul}
\usepackage{braket}
\usepackage{graphicx}% Include figure files
\usepackage{dcolumn}% Align table columns on decimal point
\usepackage{bm}% bold math
\usepackage[normalem]{ulem}
\usepackage{xcolor}
\usepackage{tikz}
\usepackage{quantikz}
\usetikzlibrary{backgrounds}
\newcommand{\kket}[1]{| #1 \rangle\!\rangle}
\newcommand{\bbra}[1]{\langle\!\langle #1 |}
\newcommand{\orcid}[1]{\href{https://orcid.org/#1}{\includegraphics[width=10pt]{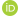}}}
\newcommand{\Proj}[1]{\big[\!\big[#1\big]\!\big]}
\newcommand{\ketbra}[2]{| #1 \rangle\!\langle #2 |}
\usepackage{scalerel}
\usepackage{hyperref}% add hypertext capabilities
\hypersetup{
     colorlinks   = true,
     linkcolor    = red,
     citecolor    = blue,
     urlcolor     = blue,
     }

\begin{document}

%\preprint{APS/123-QED}

\title{Blind-spots of Randomized Benchmarking Under Temporal Correlations}% Force line breaks with \\
 \author{Varun Srivastava$^{1}$\orcid{0000-0002-3907-5304}}
 \email{varun.srivastava@students.mq.edu.au}
 \author{Abhinash Kumar Roy$^{1}$\orcid{0000-0001-7156-1989}}
 \email{abhinash.roy@students.mq.edu.au}
 \author{Soumik Mahanti$^{1,2}$\orcid{0000-0002-0380-0324}}
 \author{Jasleen Kaur$^{1}$\orcid{0000-0002-7551-8213}}
 \author{Salini Karuvade$^{3}$\orcid{0000-0002-1513-7857}}
 \author{Alexei Gilchrist$^{1}$\orcid{0000-0003-0075-5174}}
 \email{alexei.gilchrist@mq.edu.au}
 \affiliation{$^1$Department of Physical and Mathematical Sciences, Macquarie University, Sydney NSW, Australia}
  \affiliation{$^2$School of Computer Science, University of Technology Sydney,
 Ultimo, Sydney, New South Wales 2007, Australia}
 \affiliation{$^3$School of Physics, The University of Sydney, NSW 2006, Australia}
 
\begin{abstract}
Randomized benchmarking (RB) is a widely adopted protocol for estimating the average gate fidelity in quantum hardware. However, its standard formulation relies on the assumption of temporally uncorrelated noise, an assumption often violated in current devices. In this work, we derive analytic expressions for the average sequence fidelity (ASF) in the presence of temporally correlated (non-Markovian) noise with classical memory, including cases where such correlations originate from interactions with a quantum environment. We show how the ASF can be interpreted to extract meaningful benchmarking parameters under such noise and identify classes of interaction Hamiltonians that render temporal correlations completely invisible to RB. We further provide operational criteria for witnessing temporal correlations due to quantum memory through RB experiments. Importantly, while classical correlations may remain undetectable in the ASF data, they can nonetheless significantly affect worst-case errors quantified by the diamond norm, a metric central to fault tolerant quantum computing. In particular, we demonstrate that temporal correlations may suppress worst-case errors highlighting that temporal correlations may not always have detrimental effects on gate performance. \\

\end{abstract}

\maketitle

\section{Introduction}
Quantum characterization, verification, and validation (QCVV) is essential for developing reliable quantum technologies \cite{Eisert2020,blumekohout2025QCVV}. Among QCVV tools, randomized benchmarking (RB) has become one of the most widely implemented methods for characterizing the average performance of a gate set via a single decay parameter \cite{Emerson_2005,Knill2008PRA,PhysRevLett.106.180504,PhysRevA.85.042311}.
RB offers two significant practical advantages for characterizing gates, namely, scalability and  insensitivity to state-preparation and measurement (SPAM) errors. In contrast, process tomography suffers on both counts \cite{Chuang1996_QPT}.
This practicality has driven broad adoption of RB for gate characterization across platforms including trapped-ion \cite{Knill2008PRA}, superconducting \cite{Chow2009_PRL_Superconducting,2013PRA_superconducting_two_qubit}, and neutral-atom processors \cite{Olmschenk_2010_NJP_neutral_atom,PRL_2015_neutral_atom}.

RB assumes temporally uncorrelated (i.e. Markovian) and gate-independent noise in its standard formulation. Under these assumptions, the average sequence fidelity (ASF), which is the quantity the RB protocol estimates, follows a simple exponential decay model~\cite{silva2025handsonintroductionrandomizedbenchmarking}.
Since its inception, numerous variants have been proposed to adapt the protocol to more general scenarios, namely simultaneous RB to probe multi-qubit crosstalk \cite{Gambetta2012_PRL_Sim_RB}, interleaved RB to estimate the average error of a targeted gate \cite{Magesan2012_PRL_Int_RB}, extensions beyond the Clifford group to more general gate sets \cite{Franca_2018,Helsen2019,Eisert2020}, relaxations to weak gate-dependent noise \cite{PhysRevLett.106.180504,Wallman2018randomized}, direct RB for multi-qubit native gates \cite{Proctor2019_Direct_RB} and logical-level RB on encoded qubits \cite{Combes2017_Logical_RB}. These variants largely inherit the assumption of Markovianity of the noise; however, most of the present quantum hardwares routinely exhibit temporally correlated noise \cite{McEwen2021_Google,Morris2022_IBM}. Only recently, the RB framework has been extended to accommodate non-Markovian dynamics \cite{PRXQuantum.2.040351,FigueroaRomero2022towardsgeneral}. Despite this progress, a systematic analysis of RB under diverse memory structures in the environment remains unexplored. These memory structures can be broadly categorized into classical or quantum memory depending on the type of correlations mediated by the environment \cite{Giarmatzi2021witnessingquantum,Taranto2024characterising}.

Prior studies have examined RB in the presence of classically correlated noise arising from coupling to stochastic classical fields, revealing departures from the Markovian ASF model \cite{PhysRevA.93.022303,PRXQuantum.2.040351,PhysRevA.92.022326,PhysRevA.103.022607,Veldhorst2014-db}. For instance, Ref. \cite{PhysRevA.92.022326} shows that slowly varying 1/f noise can produce decay curves better captured by multi-exponential fits. In Ref. \cite{PRXQuantum.2.040351}, two specific models, the controlled dephasing noise model and the shallow pocket model, were presented as examples of environment with classical memory, where the ASF curve showed monotonic decay. 
% Recent study of RB for the spin-boson model showed a polynomially suppressed decay of ASF in the presence of quantum memory compared to the Markovian counterpart \cite{gandhari2025quantumnonmarkoviannoiserandomized}. 
Crucially, the effective memory in the environment can be classical even when it originates from interactions with nearby quantum environment for certain classes of interaction Hamiltonians \cite{Goswami_2025}, which prominently includes couplings naturally present in the present hardware, such as ZZ interactions in superconducting devices \cite{Mundada2019_ZZ_Coupling,PRL2022_ZZcoupling}. These observations motivate a systematic study of RB in the most general classical-memory scenarios and what RB can (and cannot) reveal about both average gate errors and worst-case errors across gate sequences. Importantly, the process matrix formalism offers a unified and operational way to model and compare memory structures and assess their impact on RB \cite{PhysRevA.97.012127,Giarmatzi2021witnessingquantum,Taranto2024characterising}. 

In this work, we present a general investigation of the RB protocol in the presence of classical memory, including memory effects that originate from interactions with a quantum environment. We derive analytical expressions for the average sequence fidelity (ASF), which reveal the emergence of multiple decay parameters, and we outline how these parameters can be reliably extracted and interpreted from RB data. A central message of our study is that classical memory effects often remain hidden in ASF curves. In particular, we show that the ASF remains monotonically decreasing with sequence length under such memory. Hence, any observed non-monotonicity provides a strong diagnostic for genuinely quantum memory effects. We further identify parameter regimes in which the ASF of a classical memory process is indistinguishable from that of a Markovian model, rendering RB effectively blind to these correlations, and we specify a class of interaction Hamiltonians that lead to such RB-blindness. Finally, we demonstrate that even when RB fails to reveal memory effects, these correlations can substantially influence the worst-case performance of gates, a metric that is critical for assessing fault-tolerance thresholds \cite{Sanders_2016,PhysRevLett.117.170502}.

The paper is structured as follows: In Sec.~\ref{Randomized Benchmarking Protocol}, we provide a background for the RB protocol. We also present a brief introduction to the process matrix formalism and various memory structures in Sec.~\ref{sec:Proc.Matrix}. In Sec.~\ref{RB_process_matrix}, we cast the RB procedure in the process matrix formalism and derive the resulting ASF curves for time-dependent Markovian and classical memory scenarios. In Sec.~\ref{non-Markovian ASF and conditions for Blindness to RB protocol} we study the various blindness criteria and class of RB-blind Hamiltonians. Sec.~\ref{Worst case errors} includes discussion on worst-case error performance of gates under RB blind scenario and relevance of memory effects. We conclude and discuss future prospects in Sec.~\ref{Discussion and Conclusion}.

\section{Randomized Benchmarking Protocol}\label{Randomized Benchmarking Protocol}
In this work our analysis includes the standard randomized benchmarking protocol, where the gate set is sampled from the $n$-qubit Clifford group $\mathcal{CG}_{n}$ \cite{Ryan_2009,PhysRevLett.106.180504,Emerson_2005,PhysRevA.85.042311,PhysRevA.89.062321}. The standard RB protocol proceeds through the following steps (for other variants see \cite{Helsen_2022_General,silva2025handsonintroductionrandomizedbenchmarking} and references therein):

\begin{enumerate}
    \item {Prepare an initial state $\rho$ (ideally $\ketbra{\psi}{\psi}$).}
    
    \item The prepared state $\rho$ is subject to a sequence of $m+1$ Clifford gates $S_{i}\equiv \mathcal{C}_{m+1}^{(i)} \circ \mathcal{C}_{m}^{(i)} \circ\cdots\circ \mathcal{C}_{2}^{(i)}\circ \mathcal{C}_{1}^{(i)}$ where the first $m$ gates are sampled uniformly at random from $\mathcal{CG}_{n}$ and the final gate is a fixed \emph{motion reversal} gate chosen to invert the preceding sequence, $\mathcal{C}_{m+1}^{(i)}=\left(\mathcal{C}_{m}^{(i)} \circ\cdots\circ \mathcal{C}_{2}^{(i)}\circ \mathcal{C}_{1}^{(i)}\right)^{\dagger}$. Under ideal gate implementation $S_{i}\equiv\mathcal{I}$, i.e. the overall action is an identity map. More generally, each gate is noisy and modeled as $\mathcal{C}_{t}^{(i)}=\mathcal{G}_{t}^{(i)}\circ \mathcal{N}_{t}$, where $\mathcal{G}_{t}^{i}$ is the ideal gate implementation at time-step $t$ and $\mathcal{N}_{t}$ is a completely-positive trace-preserving (CPTP) map representing the noise present in the implementation of $\mathcal{G}_{t}^{i}$. Here, we adopt the convention that noise acts \emph{before} the ideal gate; equivalently, one can represent the noisy gate as $\mathcal{C}_{t}^{(i)}=\tilde{\mathcal{N}}_{t} \circ \mathcal{G}_{t}^{(i)}$ with appropriate modification to the analysis.

    \item {Measure with a two-outcome POVM $\{E,\,\mathbb{I}-E\}$ (ideally $E=\ketbra{\psi}{\psi}$), and estimate the sequence fidelity
    $F^{(m)}_{i} := \operatorname{Tr}\!\left[E\, S_i(\rho)\right] \in [0,1]$.}

    \item {Repeat over $k$ random sequences of the same length $m$ and average to estimate the average sequence fidelity (ASF),
    $F^{(m)} := \frac{1}{k}\sum_{i=1}^{k} F^{(m)}_{i}$. A careful discussion of finite-shot fluctuations and finite sampling over random sequences can be found in \cite{Wallman_2014,Granade_2015}. Our focus here is on the ASF; adapting those statistical analyses to non-Markovian noise is left for future work.}
    
    \item {Repeat steps 1 to 4 for several values of $m$ and fit $F^{(m)}$ as a function of $m$ to extract decay parameters and an average error rate.}
\end{enumerate}

Under the assumption of time and gate independent \emph{Markovian} noise, the ASF takes the exponential form,
\begin{equation}\label{Standard ASF}
    F^{(m)}=Aq^{m}+B,
\end{equation} 
where $q\in[0,1]$. The coefficients $A$ and $B$ capture the state preparation and measurement (SPAM) errors. 
Fitting the experimentally obtained ASF data to Eq.~\eqref{Standard ASF} yields an estimate of the decay parameter $q$, which is related to the average gate error by the relation \cite{Emerson_2005},
\begin{equation}\label{avg. gate infed.}
    r=(1-q)\frac{d-1}{d},
\end{equation}
where $d$ is the dimension of the system. A key feature of Eq.~\eqref{Standard ASF} is that it separates SPAM errors from gate errors, enabling a robust estimation of $q$ from experimental data. This separation is, in fact, a general feature of Markovian noise. For time-dependent Markovian noise, it has been shown in \cite{Wallman_2014} that the ASF takes the form
\begin{equation}\label{Mark.TimeDep.ASF}
    F^{(m)}=A \prod_{t=1}^{m}q_{t} + B,   
\end{equation}
where $q_{t} \in [0,1]$ quantifies the average gate error at time step $t$. In contrast for non-Markovian noise models, as noted in \cite{PRXQuantum.2.040351,FigueroaRomero2022towardsgeneral}, the SPAM parameters do not separate from the gate errors in general. { Moreover, the temporal correlations between noise maps makes it nontrivial to define an average error rate per gate and only allows the estimation of an average error over entire gate sequence of length $m$.}

\section{Process Matrix Formalism}\label{sec:Proc.Matrix}

\begin{figure}
    \centering
    %\includegraphics[width=1.2\columnwidth]{Process Matrix.png}
    %\include{Wfig}
    %\label{fig:enter-label}
    \includegraphics[width=0.5\textwidth]{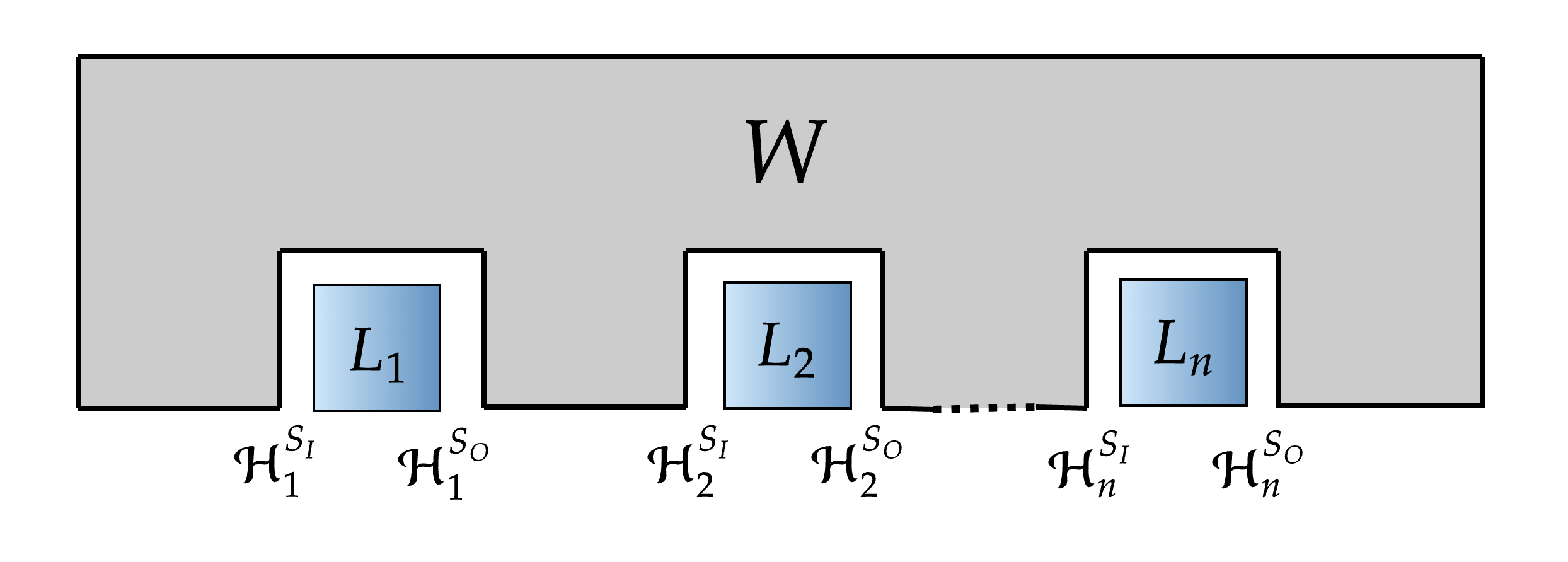}
    \caption{Schematic diagram of process matrix $W$ with $n$ intervention labs $\{L_1,L_2,\cdots,L_n\}$. The process matrix captures the most general correlations compatible with the local validity of quantum mechanics in the labs}
    \label{Fig:Proc.Mat}
\end{figure}

{In the RB protocol the system is acted on repeatedly by quantum operations (gates) at multiple time-steps, making it natural to describe the experiment using the process-matrix formalism \cite{Costa_2016,PhysRevA.97.012127, PRXQuantum.2.040351}. In a multi-time quantum process, the system is subjected to interventions at several time steps; each intervention can be viewed as a “lab” implementing a valid quantum operation, i.e., a completely positive trace non-increasing map, which in case of 
RB protocol would correspond to implementing Clifford gates followed by a final POVM measurement. Between interventions the system interacts with an environment via joint system–environment unitaries, which in our case would correspond to the noise associated to each gate implementation. Process matrices, introduced in \cite{Oreshkov2012-dk}, provide an operational description of the correlations compatible with locally valid quantum mechanics in the labs (Fig.~\ref{Fig:Proc.Mat}). When the labs are related by a definite causal order, as in the multi-time setting like RB, the process matrix satisfies causality constraints \cite{Costa_2016} and is equivalent to a quantum comb/process tensor \cite{PhysRevLett.120.040405}. This framework is widely used to model non-Markovian quantum dynamics and provides an operational notion of quantum stochastic processes \cite{PhysRevLett.120.040405,PhysRevA.97.012127}. We will use this formalism to model non-Markovian noise in RB in Sec. \ref{RB_process_matrix}.}

\begin{figure}[h]
    \centering
    \includegraphics[width=\columnwidth]{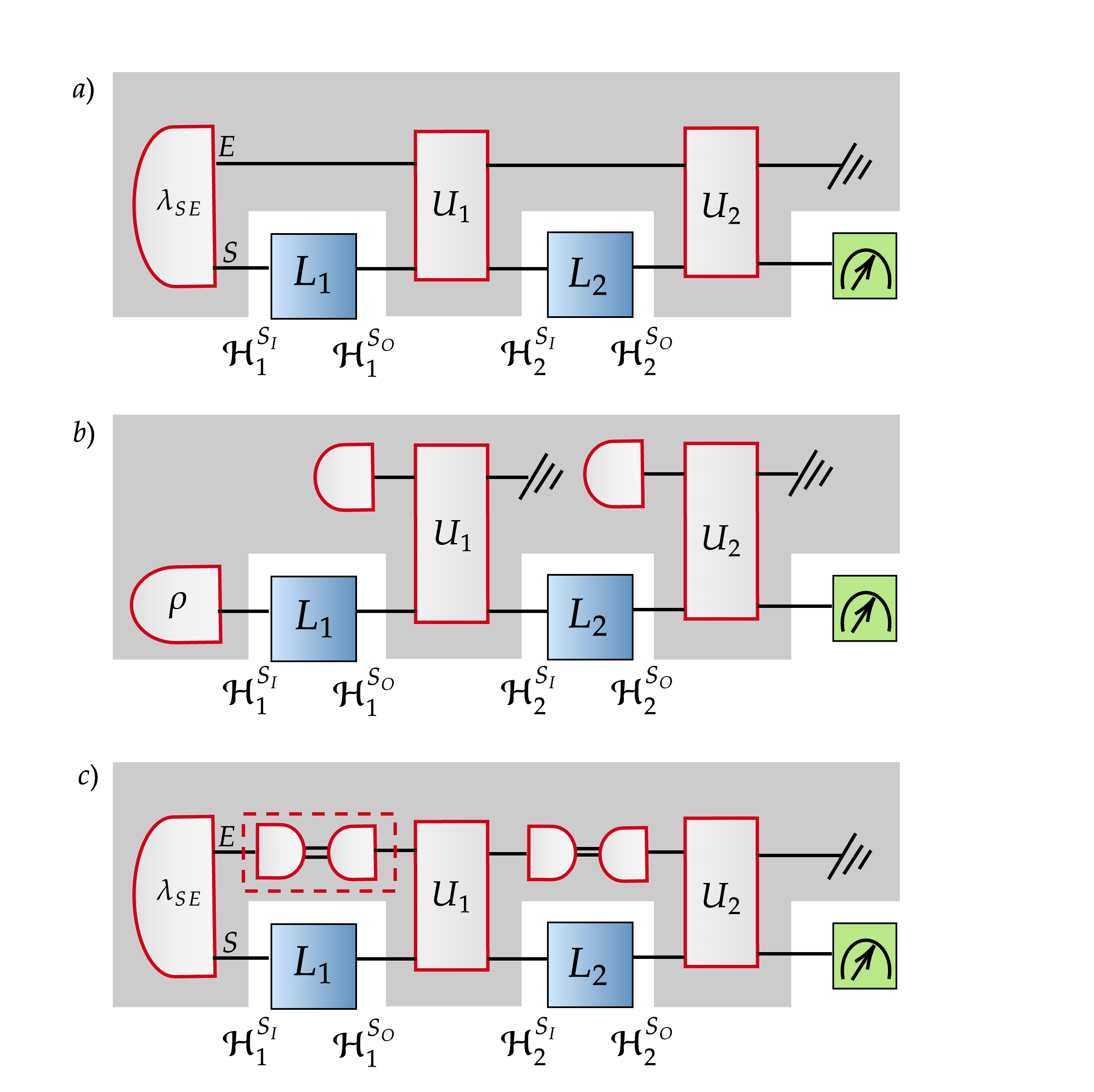}
    \caption{Schematic of two time-step multi-time process with labs $L_1,L_2$ and process matrix (in grey) capturing multi-time correlations due to system-environment interactions. 2a shows a general multi-time process with initial state and system-environment unitaries. 2b shows a Markovian process where the initial state is a product state and environment is traced out and re-prepared at each time-step. 2c represents 
    %a process with classical memory where the environment feed-forwards classical information.
    a classical memory process, where an entanglement-breaking channel acts on the environment at each step (red dashed box), preserving only classical correlations across time.}
    \label{fig:Mem.Structure}
\end{figure}

{In detail, we consider a system $S$ interacting with an environment $E$.
At each discrete time step $t$, an instrument acts locally on $S$, described by CP trace non-increasing maps
$\{\mathcal{T}_{m_t|x_t}\}:\mathcal{L}(\mathcal{H}^{S_I}_t)\to\mathcal{L}(\mathcal{H}^{S_O}_t)$,
where $x_t$ labels the instrument setting and $m_t$ the outcome. The corresponding CPTP map is
$\mathcal{T}_{x_t}=\sum_{m_t}\mathcal{T}_{m_t|x_t}$ (i.e., outcomes ignored). Between interventions, $S$ and $E$
undergo joint unitary evolution, generating correlations across time that can lead to non-Markovian behaviour
(see Fig.~\ref{fig:Mem.Structure}). For calculations we use the Choi--Jamio\l{}kowski (CJ) isomorphism \cite{CHOI1975285,JAMIOLKOWSKI1972275}.
For a linear map $\mathcal{M}:\mathcal{L}(\mathcal{H}^A)\to\mathcal{L}(\mathcal{H}^B)$, its CJ operator is
$\Proj{\mathcal{M}}=(\mathcal{I}\otimes\mathcal{M})(\kket{\mathbb{I}}\bbra{\mathbb{I}})
\in\mathcal{L}(\mathcal{H}^A\otimes\mathcal{H}^B)$, where $\kket{\mathbb{I}}=\sum_j |j\rangle\otimes|j\rangle$
is the (unnormalised) maximally entangled state in a fixed basis $\{|j\rangle\}$.}

{All multi-time system--environment correlations are then encoded in a single process matrix
$W\succeq 0 \in \mathcal{L}(\bigotimes_{t=1}^n(\mathcal{H}^{S_I}_t\otimes\mathcal{H}^{S_O}_t))$. For an $n$-step experiment, the probability of outcomes $\vec m_n=(m_1,\ldots,m_n)$ given settings
$\vec x_n=(x_1,\ldots,x_n)$ is given by the generalized Born rule \cite{Oreshkov2012-dk}}

\begin{equation}\label{gen.Born}
p(\vec m_n|\vec x_n)=\operatorname{Tr}\!\left[\,W^{T}\left(\bigotimes_{t=1}^n \Proj{\mathcal{T}_{m_t|x_t}}\right)\right],
\end{equation}
{where the transpose is taken in the CJ reference basis.
}

{$W$ can be viewed as a state up to a normalization factor, which is restricted by linear causality constraints to ensure valid probabilities for
all admissible instruments \cite{Araújo_2015}.} This “state-like” encoding of multi-time correlations is powerful, as it allows one to infer memory effects of the environment directly from the correlation structure of $W$. {Given a specific system--environment model (e.g., a sequence of joint
unitaries), $W$ can be constructed by composing CJ operators of the unitaries via the link product;} {given the CJ form of two linear} operators $\Proj{\mathcal{M}}$, $\Proj{\mathcal{N}}$, the link product $\Proj{\mathcal{M}} \star \Proj{\mathcal{N}}=\operatorname{Tr}_{{M} \cap {N}}\left[\left(\mathbb{I}^{{N}\backslash{M}} \otimes \Proj{\mathcal{M}}^{T_{{M} \cap {N}}}\right) \left(\Proj{\mathcal{N}} \otimes \mathbb{I}^{{M}\backslash{N}}\right)\right]$ where $T_{{M} \cap {N}}$ is the partial transpose over the shared Hilbert space. In Appendix~\ref{AppendA} we detail the properties of the link product and how it can be used to evaluate time-ordered process matrices and Eq.~\eqref{gen.Born}. {In the following, we provide a brief introduction to Markovian and non-Markovian processes with classical memory which would be useful to model different noise models of RB in Sec. \ref{RB_process_matrix}. For a detailed derivation of different memory structures see \cite{Giarmatzi2021witnessingquantum}.}

\subsubsection{Markovian processes}

In a Markovian process the environment does not carry information forward, so no temporal correlations are mediated between intervention times. In the process-matrix formalism this is equivalent to a product structure \cite{Costa_2016,Giarmatzi2021witnessingquantum,PhysRevA.97.012127}
\begin{equation}
\label{MarkW}
W_M=\rho^{S_I}\bigotimes_{t=1}^{n-1}\Proj{\mathcal{N}_t}^{\,S_O S_I},
\end{equation}
where $\Proj{\mathcal{N}_t}\in\mathcal{L}(\mathcal{H}^{S_O}_t\otimes\mathcal{H}^{S_I}_{t+1})$ is the CJ operator of the CPTP map from time $t$ to $t{+}1$, and $\rho^{S_I}$ is the initial system state. Operationally, this corresponds to the picture in which $S$ interacts at each step with a fresh, uncorrelated environment(see Fig.~\ref{fig:Mem.Structure}b).
Henceforth, we will ignore the superscripts on operators labeling their input/output Hilbert spaces whenever it is clear from the context.

\subsubsection{Non-Markovian processes}

Any process that does not factorize as in Eq.~\eqref{MarkW} is non-Markovian. Physically, this occurs when a shared environment interacts with the system at multiple times, retains information, and thereby mediates temporal correlations between labs. Depending on the type of correlations transmitted, non-Markovian processes can be broadly classified into \emph{classical-memory} and \emph{quantum-memory} processes \cite{Giarmatzi2021witnessingquantum,PhysRevA.110.012608,Taranto2024characterising}. In this work we focus on RB scenarios in which the noise is consistent with \emph{classical memory}.

Intuitively, classical memory means that the environment effectively generates and stores a \emph{classical record} which is fed forward to influence later dynamics (see Fig.\ref{fig:Mem.Structure}c). We call this the \emph{classical feed-forward (CFF)} mechanism: between times $t$ and $t{+}1$ the effective environmental action is a conditional instrument
$\{\mathcal{N}_{a_t|x_t}\}_{a_t}$ whose choice of setting $x_t$ may depend on the classical history $(\vec a_{t-1},\vec x_{t-1})$ via
$p(x_t|\vec a_{t-1},\vec x_{t-1})$. The resulting process matrix is
\begin{equation}
\label{W_CFF}
W_{CFF}=\sum_{\vec{x}_n,\vec{a}_n}p(x_{0})\,\rho_{a_0|x_0}\bigotimes_{t=1}^{n-1}
p(x_{t}|\vec{a}_{t-1},\vec{x}_{t-1})\Proj{\mathcal{N}_{a_{t}|x_t}},
\end{equation}
where $\{\rho_{a_0|x_0}\}_{a_0}\in\mathcal{L}(\mathcal{H}_{1}^{S_I})$ are sub-normalized states satisfying
$\operatorname{Tr}\!\left(\sum_{a_0}\rho_{a_0|x_0}\right)=1$, i.e for each setting $x_0$, the state $\rho_{a_0|x_0}$ is selected with probability $\operatorname{Tr}(\rho_{a_0|x_0})$.

A useful special case is \emph{classical common-cause (CCC)}, where the environment is characterized by a single classical label $x$ and applies deterministic CPTP maps conditioned on $x$:
\begin{equation}
\label{W_CCC}
W_{CCC}=\sum_{x} p(x)\,\rho_{x}\otimes \Proj{\mathcal{N}_x}\otimes \underbrace{\cdots}_{n-1 ~\text{times}}\otimes\Proj{\mathcal{N}_x}.
\end{equation}
CCC processes are therefore convex mixtures of Markovian processes in Eq.~\eqref{MarkW}.

Finally, such classical correlations can arise either from coupling to classical stochastic fields or from genuine interactions with a quantum environment \cite{Goswami_2025,White2025whatcanunitary}; importantly, any classical-memory process can be \emph{simulated} by a classical stochastic model. In Appendix~\ref{AppenB} we summarise system--environment Hamiltonian mechanisms known to generate these effects. Any process matrix that is neither Markovian nor of classical-memory type constitutes a \emph{quantum-memory} process.
{In RB, the process matrix provides a compact description of the underlying noise. For standard RB with Markovian noise, $W$ factorizes as in Eq.~\eqref{MarkW}. Here, instead, we focus on classical-memory noise models described by Eqs.~\eqref{W_CFF} and~\eqref{W_CCC}, with particular emphasis on the CCC class.}

\section{Modeling non-Markovian noise in Randomized Benchmarking} \label{RB_process_matrix}
\begin{figure}[h]
    \centering
    \includegraphics[width=\columnwidth]{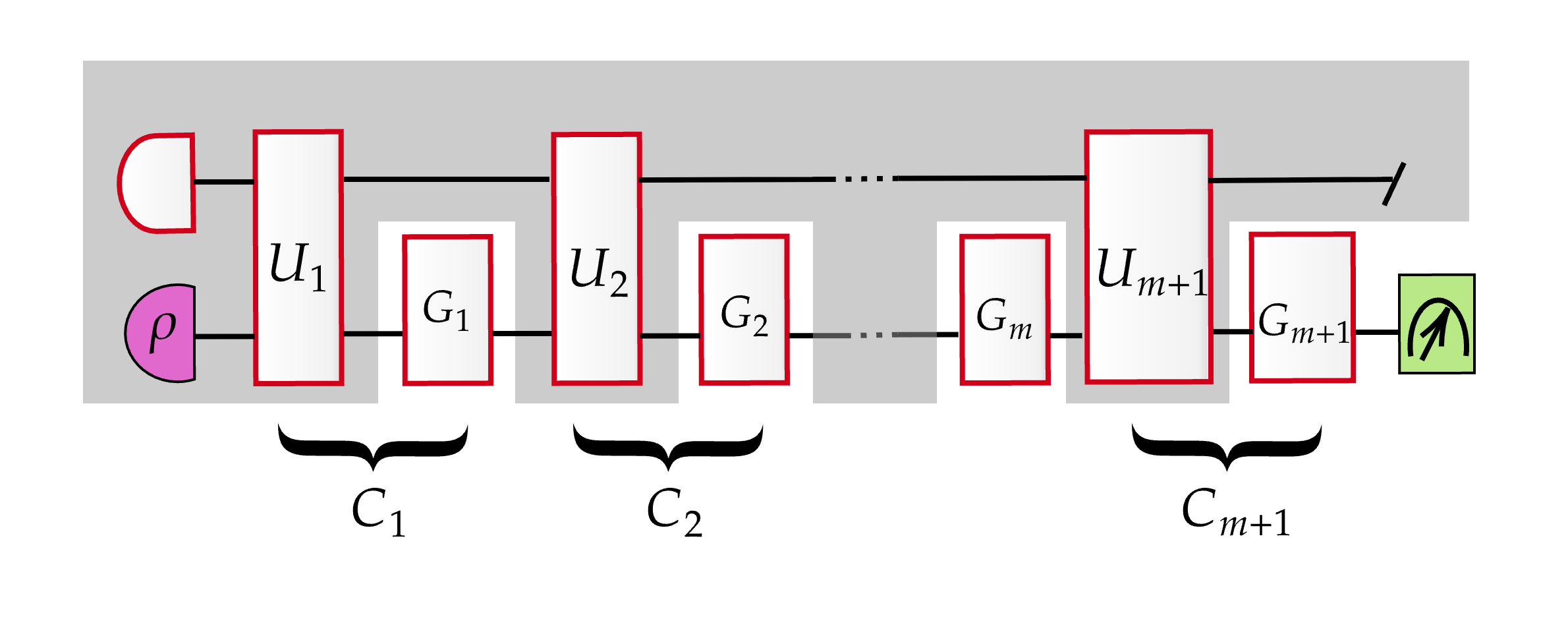}
    \caption{\emph{RB protocol in the presence of non-Markovian noise}: The process matrix construction provides a natural framework to study non-Markovian effects in RB. Each noisy gate $\{\mathcal{C}_1, \mathcal{C}_2, \cdots,\mathcal{C}_{m+1}\}$ can be divided into ideal Clifford gates $\{G_1,G_2,\cdots,G_{m+1}\}$ which act only on the system and the noise is captured by the process matrix comb (in grey).}
    \label{fig:RB_protocol}
\end{figure}

In this section, we demonstrate how the process matrix formalism can be applied to model non-Markovian noise within the randomized benchmarking protocol. Consider a model with initially uncorrelated system-environment state $\rho^S \otimes \sigma^E$.
The action of the noisy gate is given by {$\mathcal{C}_{t} = \left(\mathcal{I} \otimes \mathcal{G}_{t}\right)\circ\mathcal{U}^{SE}_{t}$, where $\mathcal{U}^{SE}_{t}$} is the joint system-environment unitary and $\mathcal{G}_{t}$ are the ideal gates sampled from the $n$-qubit Clifford group. In the final time-step we apply the motion-reversal gate and measure with POVM element $E$. All components except for the control operations $\{\mathcal{G}_{t}\}_{t=1}^{m+1}$ and $E$ can be collected into the process matrix, which acts as a higher-order quantum operation mapping the chosen sequence of gates to corresponding sequence fidelity (see Fig.~\ref{fig:RB_protocol}) \cite{taranto2025higherorderquantumoperations}. More generally, if the initial state is correlated, one can apply a trace-and-replace operation on $S$ to prepare an effective product state. Operationally, this requires the initial $S$--$E$ coupling (and hence initial correlations) to be weak. In the following subsections, we first formulate the ASF calculation within the process matrix framework, which allows us to model Markovian as well as non-Markovian gate errors with classical memory.

\subsection{ASF calculation using process matrix formalism} \label{ASF calculation using process matrix formalism}

{The instruments in the RB protocol are unitary gates $\{\mathcal{G}_t\}\in \mathcal{CF}_n$ (see Fig. \ref{fig:RB_protocol}). For a sequence $S_\alpha$, the operation at time $t$ is the unitary channel $\mathcal{G}_t^{(\alpha)}(\cdot)=G_t^{(\alpha)}(\cdot)G_t^{(\alpha)\dagger}$ represented in the CJ form as $\Proj{\mathcal{G}_t}$. Using the link product to concatenate the initial system-environment state and joint unitaries, we obtain the process matrix $W$. The sequence fidelity can be expressed as the link product between $W$ and the instruments as the following}
\begin{equation}
    \label{eq:3letter_eqn}
    F^{(m)}_\alpha = W \star \mathbb{T}_\alpha,
\end{equation}
where
\begin{equation}
\label{W_Talpha_defs}
\begin{aligned}
W &= (\rho \otimes \sigma) \star \Proj{\mathcal{U}^{SE}_1} \star \cdots \star \Proj{\mathcal{U}^{SE}_{m+1}} \star \mathbb{I}^{E},\\
\mathbb{T}_\alpha &= \Proj{\mathcal{G}^{(\alpha)}_1} \star \cdots \star \Proj{\mathcal{G}^{(\alpha)}_m} \star \Proj{\mathcal{G}^{(\alpha)}_{m+1}} \star E^T .
\end{aligned}
\end{equation}
Moreover, we define a set of new Clifford operations, $\mathcal{K}_{1}^{(\alpha)}=\mathcal{G}_{1}^{(\alpha)}$ and $\mathcal{K}_{i}^{(\alpha)}=\mathcal{G}_{i}^{(\alpha)} \circ \mathcal{K}_{i-1}^{(\alpha)}~ ~\forall~ i \geq 2$. Since the set $\mathcal{CF}_{n}$ forms a group, $\mathcal{K}_{i}^{(\alpha)}\in \mathcal{CF}_{n}~~ \forall~ i$, using the cyclicity and the linearity of trace operation, Eq.~\eqref{eq:3letter_eqn}  can be equivalently represented as 
\begin{equation}
    \label{SeqFid.ProcMat2}
    F_{\alpha}^{(m)}=\tilde{W}_{\alpha} \star {\Gamma},
\end{equation}
where $\Gamma=\kket{\mathbb{I}}\bbra{\mathbb{I}}^{\otimes (m)} \otimes E^T$. { Note that $\kket{\mathbb{I}}\bbra{\mathbb{I}}= \Proj{\mathcal{I}}$, is the Choi operator of the identity channel. Hence, after this re-writing, the new instruments applied to process is simply $m$ identity channel slots together with a final POVM measurement and all dependence on the particular RB sequence $\alpha$ is absorbed into an effective, sequence-dependent process matrix $\tilde{W}_{\alpha}$, given by,} 

\begin{equation}
\label{tildeW_short}
\begin{aligned}
\tilde{W}_{\alpha}&=\mathbb{K}^{(\alpha)}\,W\,\mathbb{K}^{\dagger(\alpha)},\\
\mathbb{K}^{(\alpha)}
&=\mathbb{I}~\!\bigotimes_{t=1}^{m}\!\left({K}_{t}^{T(\alpha)} \otimes {K}_{t}^{\dagger(\alpha)} \right),
\end{aligned}
\end{equation}
where superscript $T$ denotes the transpose in the specified basis. We summarize the above in the following theorem.

%{Equation~\eqref{SeqFid.ProcMat2} thus shows that the dependence on the RB sequence $\alpha$
%can be absorbed into a sequence-induced transformation $W \mapsto \tilde{W}_{\alpha}$,
%while the instruments acting on $\tilde{W}_{\alpha}$ reduces to $m$ identity-channel slots together with
%the final POVM element $E$.}

{
\begin{theorem}\label{thm1}
    Given an initial system environment state $\rho \otimes \sigma \in  \mathcal{L}(\mathcal{H}^{S}_{0}\otimes\mathcal{H}^{E}_{0})$ and a sequence of noisy gates, the noise structure is captured by the process matrix $W$ defined in Eq.~\eqref{W_Talpha_defs}. The ASF is 
    \begin{equation}
    \label{avg.seq.fidelity_short}
    \bar{F}^{(m)}=\frac{1}{\Omega^m}\sum_{\alpha=1}^{\Omega^m}\tilde{W}_{\alpha} \star {\Gamma},
    \quad \Omega=|\mathcal{CF}_n|.
    \end{equation}
\end{theorem}
}
The details of the derivation is provided in  Appendix \ref{addn.appen.ASF}. { We  note that in \cite{PRXQuantum.2.040351}, the authors also provided a general expression for ASF for an arbitrary process matrix. However, the ASF is formulated at a level of generality that does not
directly translate into experimentally practical procedures for extracting relevant metrics of interest unless we specify details of the noise model. In our work we focus on a physically
well-motivated subclass of non-Markovian processes (classical-memory models) for which the ASF acquires an
interpretable structure and the relevant parameters can be estimated from RB data.}

{We conclude this section by showing that Eq.~\eqref{avg.seq.fidelity_short} reproduces the standard ASF for a time-dependent Markovian process whose process matrix has the product form $W_M=\rho \star \Proj{\mathcal{N}_1} \star \cdots \star \Proj{\mathcal{N}_{m+1}}$, where $\rho$ is the initial system state and $\Proj{\mathcal{N}_t}$ denotes the Choi operator of the CPTP noise map at time step $t$. With our convention that noise acts before the ideal Clifford gate (Fig.~\ref{fig:RB_protocol}), we define $\tilde{\rho}:=\rho \star \Proj{\mathcal{N}_1}=\mathcal{N}_1(\rho)$, so that $W_M$ can be written as $W_M=\tilde{\rho}\otimes \Proj{\mathcal{N}_2}\otimes\cdots\otimes \Proj{\mathcal{N}_{m+1}}$.}

{Applying the sequence-induced transformation $W\mapsto \tilde{W}_\alpha$ from Eq.~\eqref{tildeW_short} and substituting into Eq.~\eqref{avg.seq.fidelity_short}, the ASF reduces to}
\begin{equation}
\label{modified_mark_ASF}
\bar{F}^{(m)}_M=\left(\tilde{\rho}\otimes Z_2 \otimes \cdots \otimes Z_{m+1}\right)\star \Gamma,
\end{equation}
where $Z_t$ is the twirl of $\Proj{\mathcal{N}_t}$ over $\mathcal{CF}_n$, namely
$Z_t:=\frac{1}{\Omega}\sum_{K_t\in\mathcal{CF}_n}\bigl(K_{t}^{T(\alpha)} \otimes K_{t}^{\dagger(\alpha)}\bigr)\,\Proj{\mathcal{N}_t}\,\bigl(K_{t}^{T(\alpha)} \otimes K_{t}^{\dagger(\alpha)}\bigr)^{\dagger}$, with $\Omega=|\mathcal{CF}_n|$. Since $\mathcal{CF}_n$ forms a unitary $2$-design, this average is equal to the Haar second moment. Using Schur--Weyl duality (see Appendix~\ref{addn.appen.ASF} for details), the twirl projects onto $\mathrm{span}\{\mathbb{I},\kket{\mathbb{I}}\bbra{\mathbb{I}}\}$, yielding
\[
Z_t= c_{1}(\mathcal{N}_t)\,\frac{\mathbb{I}}{d} \;+\; c_{2}(\mathcal{N}_t)\,\kket{\mathbb{I}}\bbra{\mathbb{I}},
\]
with coefficients
\begin{equation}\label{2nd mom coeff}
\begin{aligned}
c_{1}(\mathcal{N}_t) &= \frac{d\,\operatorname{Tr}\!\left(\Proj{\mathcal{N}_t}\right) -  \operatorname{Tr}\!\left(\kket{\mathbb{I}}\bbra{\mathbb{I}}\, \Proj{\mathcal{N}_t}\right)}{d^2 - 1}, \\
c_{2}(\mathcal{N}_t) &= \frac{\operatorname{Tr}\!\left(\kket{\mathbb{I}}\bbra{\mathbb{I}}\, \Proj{\mathcal{N}_t}\right) - d^{-1}\operatorname{Tr}\!\left(\Proj{\mathcal{N}_t}\right)}{d^2 - 1}.
\end{aligned}
\end{equation}
Substituting this form of $Z_t$ into Eq.~\eqref{modified_mark_ASF} gives the desired time-dependent Markovian ASF,
$\bar{F}^{(m)}_M=A\prod_{t=1}^{m} q_t + B$, as in Eq.~\eqref{Mark.TimeDep.ASF}, where $A=\operatorname{Tr}\!\left[E\left(\mathcal{N}_{1}(\rho)-\mathbb{I}/d\right)\right]$, $B=\frac{1}{d}\operatorname{Tr}(E)$, and $q_t=c_2(\mathcal{N}_t)$.

\subsection{Non-Markovian processes with classical memory}\label{Non-Markovian processes with classical memory}
We now derive the ASF for non-Markovian noise models whose multi-time correlations exhibit classical memory. In the following, we analyze both cases for CCC and CFF type memory models and derive the modified ASF expression from RB protocol.

\subsubsection{ASF for classical common cause}

The process matrix for a CCC-type process is given by Eq.~\eqref{W_CCC} and represents a \emph{convex mixture of Markovian processes}. Accordingly, each run of the RB protocol may be viewed as sampling a noisy Markovian branch from an associated probability distribution. This class of noise models captures several experimentally relevant scenarios, including drifting qubit resonance frequencies \cite{PhysRevA.92.022326}, stochastic electromagnetic field fluctuations, and \(ZZ\)-type Hamiltonian couplings between qubits in superconducting architectures \cite{Goswami_2025}. Therefore, despite its apparent simplicity, this noise model is of considerable practical relevance. Moreover, in \cite{Goswami_2025}, a class of system environment interaction Hamiltonians were identified which give rise to such memory models and we summarize the results in Appendix~\ref{AppenB}. 

We can write Eq.~\eqref{W_CCC} as $W_{CCC}=\sum_{x}p_{x} W^{(x)}_{M}$ where each $W^{(x)}_{M}$ corresponds to a Markovian process matrix with associated noise map $\mathcal{N}_x$ and weight factor $p_x$. Here we are considering the scenario, where each Markovian process matrix $W^{(x)}_{M}$ correspond to time-independent process. It is straightforward to extend the analysis to convex mixture of time-dependent Markovian processes.
From Eq.~\eqref{SeqFid.ProcMat2} we get the sequence fidelity $F_{\alpha}^{(m)}=\sum_{x} p_x \tilde{W}^{(x)}_{M,(\alpha)} \star \Gamma$, for some sequence of gates $S_{\alpha}$. Using Eq.~\eqref{avg.seq.fidelity_short} we obtain the ASF
\begin{equation}
    \label{ASF_CCC}
    \bar{F}^{(m)}=\sum_{x} p_{x}A_{x}q_{x}^{m} + B,
\end{equation}
where $A_{x}=\operatorname{Tr}\left(E\left(\mathcal{N}_{x}(\rho)-\mathbb{I}/d\right)\right)$, $B=\operatorname{Tr}(E)/d$ and $q_x=c_2(\mathcal{N}_x)$ as defined in Eq.~\eqref{2nd mom coeff}.

{We thus obtain an ASF that is a \emph{sum of exponentials} rather than a single exponential, therefore estimating the decay parameters becomes a multi-exponential fitting problem. Such problems are well studied in signal processing; see \cite{Helsen_2022_General,Roy1986-ck,li2022stablesuperresolutionlimitsmallest} and Sec.~\ref{Complete blindness to RB}. We also note that the SPAM coefficient $A_x=\operatorname{Tr}\left(E\left(\mathcal{N}_{x}(\rho)-\mathbb{I}/d\right)\right)$ is no longer independent of the decay parameters. The origin of this coupling is that, in standard RB, the first noise map is not twirled in the same way as the subsequent steps. In the Markovian setting this does not cause difficulties because noise at different times is independent, so preparation errors cannot influence later dynamics beyond their effect on the initial state. In the presence of temporal correlations, however, the environment can correlate the initial preparation with subsequent noise, so SPAM contributions can become coupled to the effective decay behaviour. Consequently, even for the CCC model, the standard RB analysis must be modified to properly account for correlated SPAM errors.}

{\emph{Randomization of the initial state and final measurement:}
To decouple SPAM from the decay parameters, we additionally twirl the first noise map by repeating the RB experiment over a family of input states and matched two-outcome measurements. Concretely, we consider preparations $\rho=\ketbra{\psi}{\psi}$ together with POVMs $\{E,\mathbb{I}-E\}$ where  $E=\ketbra{\psi}{\psi}$. }

{We implement this by sampling a unitary $\mathcal{U}$ (ideally from the Haar measure, or equivalently from a unitary $2$-design) and preparing and measuring in the rotated basis. Assuming that the SPAM procedures are uncorrelated with the subsequent RB dynamics, we model the implemented preparation and measurement as}
\begin{equation}
\label{randomSPAM}
\rho_{\mathcal{U}}=\mathcal{U}\circ \mathcal{P}\!\left(\ketbra{0}{0}\right),
\qquad
E_{\mathcal{U}}=\mathcal{U} \circ \mathcal{M}\!\left(\ketbra{0}{0}\right),
\end{equation}
{where $\mathcal{P}$ and $\mathcal{M}$ are CPTP maps that capture the errors in the preparation and measurement stages, respectively, and $\mathcal{U}$ is the randomizing unitary. This additional randomization effectively twirls the first noise map. Substituting Eq.~\eqref{randomSPAM} in the expression for $A_x$ and $B$ we obtain, 
}
\begin{equation}
    \label{mod.SPAMS}
    \begin{aligned}
    A_x&=q_x \operatorname{Tr}\left(\mathcal{M}(\ketbra{0}{0}) \left[\mathcal{P}(\ketbra{0}{0})-\frac{\mathbb{I}}{d}\right]\right)\\
    &=q_x~A,\\
    B &= \frac{1}{d}.
    \end{aligned}
\end{equation}
Thus we observe that the SPAM errors have been effectively decoupled and we get the ASF for CCC models below.

{\begin{theorem}
    For a non-Markovian CCC type error model, the modified RB protocol under randomisation of initial states and final measurements is 
    \begin{equation}
    \label{Eff.Mod.ASF.CCC}
    \bar{F}^{(m)}_{CCC}= A \sum_{x} p_{x} q_{x}^{m+1}+B,
\end{equation}
where now $A$ encodes the SPAM errors and $B$ is determined by the dimension of the system. 
\end{theorem}}
Note that this procedure would require more resources in terms of the number of runs of the RB experiment, however this is expected as we are trying to benchmark correlated noise.

\emph{Average error rates:}
Assuming that Eq.~\eqref{Eff.Mod.ASF.CCC} can in principle be fitted efficiently to extract the parameters $\{p_x,q_x\}$, the interpretation of the ASF differs significantly from the Markovian case. Although the error maps $\mathcal{N}_x$ are CPTP maps acting only on the system, they are correlated through a common environmental degree of freedom. For Markovian errors, the independence of error maps allows one to assign an average error rate to each gate individually. In contrast, for errors modeled by CCC, it is not meaningful to define a per-gate error rate. Rather we can provide the average error rate of a \emph{sequence} of gates of length $m$. For a given gate sequence $S_i$ of length $m$, the entire process is described by the CPTP map $\mathcal{R}_{S_i}$ such that $\mathcal{R}_{S_{i}}(\rho)=\sum_x p_{x} \mathcal{G}_{m+1}\circ \mathcal{N}_{x}\cdots\circ\mathcal{G}_{1}\circ \mathcal{N}_{x}(\rho)$. With ideal state preparation and measurement ($\mathcal{P}=\mathcal{M}=\mathcal{I}$),  the average fidelity of $\mathcal{R}_{S_i}$ is $\hat{F}^{(m)}_{(\mathcal{R}_{S_i})}$ and its average error rate is $\hat{r}_{{(\mathcal{R}_{S_i})}}=1-\hat{F}^{(m)}_{(\mathcal{R}_{S_i})}$. Averaging over all sequence $S_i$ gives the average error rate of the $m$ length process.

{\begin{corollary}
\label{avg.errors.coro}
For general non-Markovian noise, it is not, in general, meaningful to assign a sequence-independent ``average error per gate.'' Instead, one should quantify performance at the level of the implemented circuit depth. In particular, for CCC noise models the average error rate for a sequence of length $m$ is
\begin{equation}
\label{avg.error.rateCCC}
r(m)=\frac{d-1}{d}\left(1-\sum_{x} p_{x}q_{x}^{m+1}\right).
\end{equation}
\end{corollary}}

Thus, the parameters $\{p_x,q_x\}$ provide a quantitative measure of how far, on average, the system state deviates from the ideal target state after circuit depth $m$. Even though per-gate error rates are ill-defined, the framework still yields a meaningful characterization of overall computational accuracy for sequences of arbitrary length.

\emph{An example of multiple exponential fitting:}
Ref. \cite{Helsen_2022_General} shows that several RB variants under Markovian noise can yield ASFs that are sums of (matrix) exponentials and studies how to extract multiple decay parameters from such data. In particular, the authors use high-resolution spectral estimation methods such as ESPRIT \cite{Roy1986-ck} and MUSIC \cite{1143830}, and identify regimes where reliable estimation is possible. Since the CCC ASF derived here is likewise a sum of exponentials, the same techniques apply. Below we illustrate this by extracting multiple decay parameters for a simple CCC model using ESPRIT.

Consider a qubit process where the system-environment interaction unitary at every time step is the controlled unitary of the form $U^{ES} =\ketbra{0}{0}^{E}\otimes U_1 + \ketbra{1}{1}^{E}\otimes U_2$, where $U_i = e^{-i\lambda_i}\ketbra{\lambda_i}{\lambda_i}+ e^{-i\lambda_i^{\perp}}\ketbra{\lambda^{\perp}_i}{\lambda^{\perp}_i}$ for $i\in\{1,2\}$, $|\lambda_i\rangle$  and $|\lambda_i^{\perp}\rangle$ represents an arbitrary vector and the corresponding orthogonal vector, respectively in the Hilbert space of the system, and $\lambda_i, \lambda_i^{\perp}$ can take real values. This interaction results in a CCC process, which is a convex sum of two Markovian processes, where the weights depend on the initial environment state (Appendix~\ref{AppenB}). If the initial environment state is an equal superposition of $|0\rangle$ and $|1\rangle$, the weights will be identical and the resulting ASF will be of the form $A(q_1^{m} + q_2^{m})/2+B$, where the decay parameters $q_i = (4\cos^2(\Delta\lambda_i/2)-1)/3$, with $\Delta\lambda_i = \lambda_i -\lambda_i^{\perp}$ for $i\in\{1,2\}$. Using this model, we simulate the randomized benchmarking decay curve for parameters \( q_1 = 0.9 \) and \( q_2 = 0.99 \). The RB protocol is implemented with 300 random Clifford sequences, each measured over 5000 shots, and sequence lengths ranging from 5 to 200. We first fit the simulated data using a single-exponential decay model, corresponding to the traditional RB fitting approach. Employing the \texttt{scipy.optimize} module in Python, we obtain the fitted parameters \( q = 0.981 \), \( A = 0.303 \), and \( B = 0.538 \).

\begin{figure}
    \centering
    \includegraphics[width=1\linewidth]{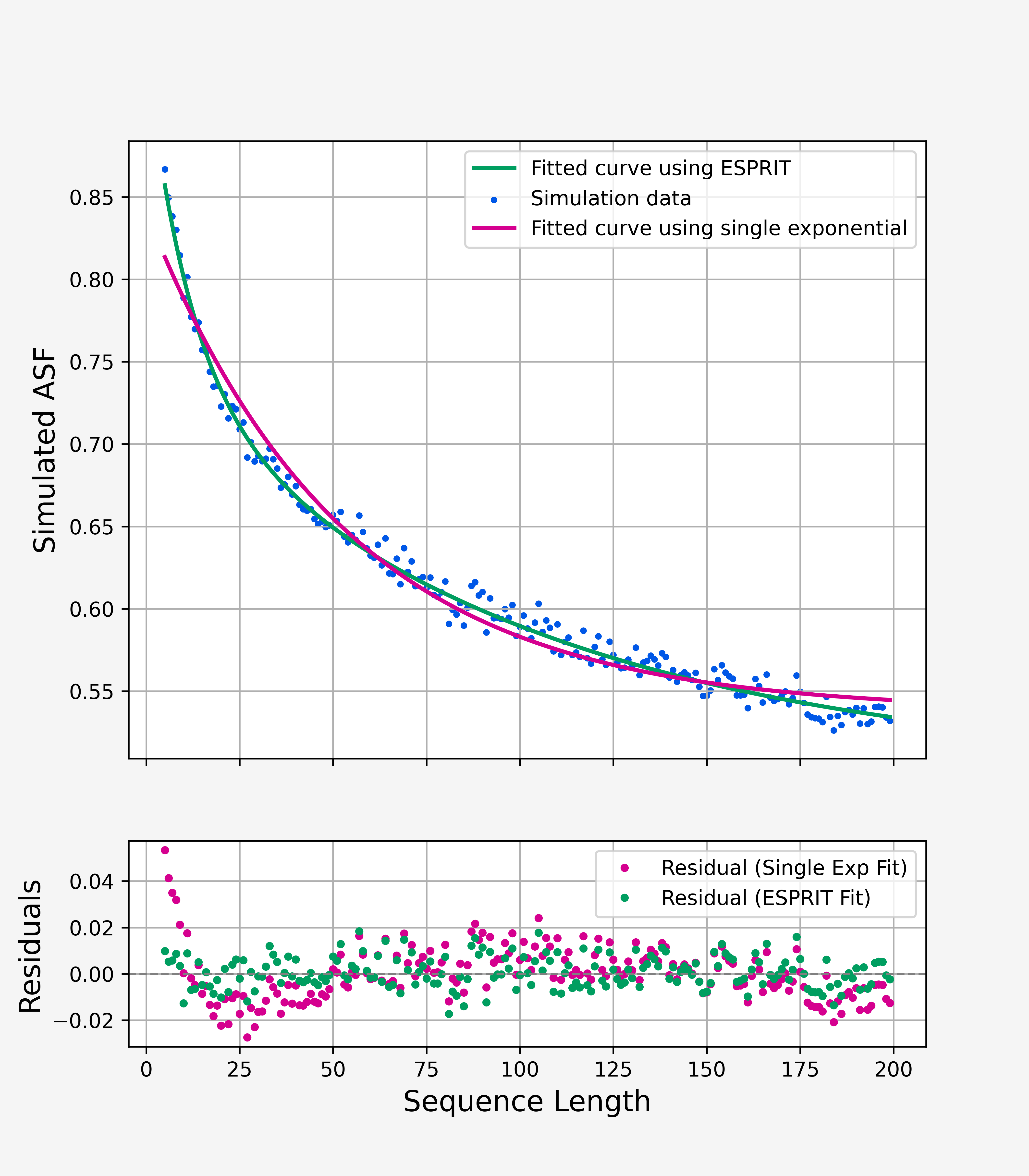}
    \caption{Comparing the performance of ESPRIT with single exponential fit for the simulated data with exponents 0.9 and 0.99. The RMSE (root mean square error) and adjusted $R^2$ values for ESPRIT fit is 0.0068 and 0.9908 respectively, while the same parameters take respective values 0.0116 and 0.9735 for single exponential fitting.}
    \label{Single and double fit}
\end{figure}

\begin{figure}[h!]
    \centering
    \includegraphics[width=1\linewidth]{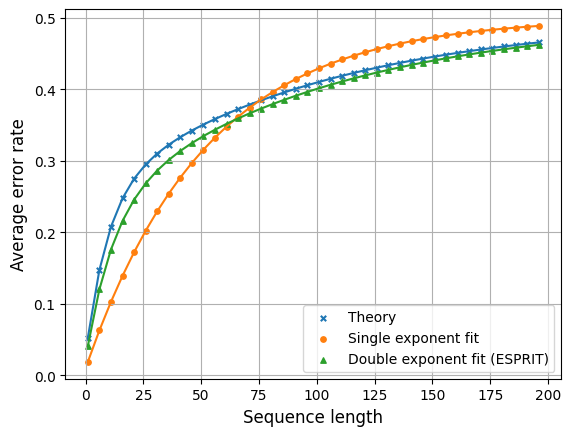}
    \caption{\emph{Average error rate versus sequence length.} Comparison of the average error rates predicted by a single-exponential fit with parameter $q=0.981$ and by a double-exponential model obtained using ESPRIT, with parameters $q_1=0.918$ and $q_2=0.990$. {The single-exponential curve underestimates the average error rate at short sequence lengths, where the ASF decay is dominated by the parameter $q_1$. Conversely, at large sequence lengths, the decay is governed primarily by $q_2$. The ESPRIT fit (green) also underestimates the theoretical curve (blue), which is due to the sampling of only 300 random Clifford sequences. Nevertheless, the ESPRIT fit provides a clearly improved description of the data.}}
    \label{Average error rates}
\end{figure}

Next, we apply the ESPRIT method (see Appendix~\ref{AppenC} for details on the fitting procedure) to extract two decay components, yielding \( q_1 = 0.918 \), \( q_2 = 0.990 \), and corresponding weights \( p_1 = 0.451 \) and \( p_2 = 0.549 \). The associated SPAM parameters are found to be \( A = 0.435 \) and \( B = 0.503 \) (see Fig.~\ref{Single and double fit}).
Finally, using Eq.~(\ref{avg.error.rateCCC}), we compare the average error rates derived from the theoretical model with those obtained from the single-exponential and ESPRIT fits, as illustrated in Fig.~\ref{Average error rates}. {This example illustrates that the multiple exponent fit parameters provide a physically meaningful interpretation in terms of average errors as defined in Corollary \ref{avg.errors.coro} and Eq.~\eqref{avg.error.rateCCC} provides an interpretation of these parameters in presence of CCC type errors.}

\subsubsection{ASF for classical feed-forward (CFF)}\label{ASF for classical feed-forward (CFF)}

{The classical feed-forward (CFF) model generalizes CCC and represents the most general form of \emph{classical} memory in our framework. The central idea is that the environment carries a hidden classical state that can change during the RB sequence. At each time step $t$, the noise acting on the system is determined by the current value of this classical state. The interaction also produces a classical outcome, which is then used to update the classical state for subsequent steps. As a result, the effective noise at later times may depend on earlier times, but only through this classical record.}

{Formally, each system--environment interaction at time $t$ is described by a quantum instrument. The instrument applied at time $t$ is selected according to a conditional probability distribution that depends on the past classical record (previous instrument settings and outcomes).}

The process matrix for the RB setup in Fig.~\ref{fig:RB_protocol} can be written as $$W_{CFF}=\sum_{\vec{x}_n,\vec{a}_n}p(x_{1})\,\rho_{a_1|x_1}\bigotimes_{t=2}^{m+1}
p(x_{t}|\vec{a}_{t-1},\vec{x}_{t-1})\Proj{\mathcal{N}_{a_{t}|x_t}},$$
where for the initial state $\rho=\mathcal{U}\circ\mathcal{P}\ketbra{0}{0}$ we define $\rho_{a_1|x_1}=\mathcal{N}_{a_1|x_1}(\rho)$ and for time-step $t$, $\mathcal{N}_{a_t|x_t}$ is a completely positive linear map such that $\sum_{a_t}=\mathcal{N}_{a_t|x_t}$ is a CPTP map. The labels $\vec{x}_t\equiv \{x_1,\cdots,x_t\}$ are the instrument settings and $\vec{a}_t\equiv \{a_1,\cdots,a_t\}$ are the measurement outcomes of the instruments.

Employing the randomization of initial states and final measurements discussed in previous subsection with SPAM procedures are described by Eq.~\eqref{randomSPAM}, the ASF for CFF processes take the following functional form.

\begin{theorem}
{The ASF for classical feed-forward model after applying the randomization of initial states and final measurements is}
\begin{equation}
    \label{ASF feed-forward}
    \bar{F}^{(m)}_{CFF}=A \left(\sum_{\vec{a},\vec{x}} \prod_{i=0}^{m} \gamma_{a_i,x_{i}|\vec{x}_{i-1},\vec{a}_{i-1}}\right)+B
\end{equation}
where $A=\operatorname{Tr}\left(\mathcal{M}(\ketbra{0}{0}) \left[\mathcal{P}(\ketbra{0}{0})-\mathbb{I}/d\right]\right)$, $B=1/d$ and $\gamma_{a_i,x_{i}|\vec{a}_{i-1},\vec{x}_{i-1}}=p(x_{i}|\vec{a}_{i-1},\vec{x}_{i-1}) \beta_{a_i|x_i}$ such that 
\begin{equation}
    \label{beta value}
    \beta_{a_{i}|x_i}=\frac{1}{d^{2}-1}\left[\operatorname{Tr}\left(\Proj{\mathcal{N}_{a_{i}|x_i}}\kket{\mathbb{I}}\bbra{\mathbb{I}}\right)-d^{-1}\operatorname{Tr}\left(\Proj{\mathcal{N}_{a_{i}|x_i}}\right)\right].
\end{equation}
\textrm{Note that here} $\sum_{a_{i}} \operatorname{Tr}\left(\Proj{\mathcal{N}_{a_{i}|x_i}}\right)=d$.
\end{theorem}

{The decay parameters in Eq.~\eqref{ASF feed-forward} allow one to define an average error rate for length-$m$ sequences, in direct analogy with Eq.~\eqref{avg.error.rateCCC}. In practice, however, reliably extracting these parameters depends on the specifics of the underlying interaction model and on how well separated the decay modes are, since near-degenerate parameters are difficult to resolve with algorithms like ESPIRIT. Nevertheless, Eq.~\eqref{ASF feed-forward} provides a principled interpretation of RB outcomes in the presence of CFF-type noise, and a framework for benchmarking devices when such effects are suspected. Moreover, as we will show in the next section, this discussion of the most general classical memory noise is important for the witness for quantum non-Markovianity (See Corollary \ref{Cor:QM_witness})}.

\section{non-Markovian ASF and conditions for Blindness to RB protocol}\label{non-Markovian ASF and conditions for Blindness to RB protocol}
The RB protocol estimates the average error rate of a gate set without incurring the resource costs of full process tomography \cite{White_PRXQ_2022,roy2026practicaltomographymultitimeprocesses}. However, this averaging procedure can also mask important non-Markovian effects. As shown in Sec.~\ref{Non-Markovian processes with classical memory}, classical-memory error models can produce ASFs with multiple decay parameters, which can in principle reveal signatures of non-Markovian behavior. Nevertheless, there exist scenarios where the resulting ASF is indistinguishable from that of a purely Markovian noise model.

In this section, we establish criteria under which classical-memory errors are effectively \emph{blind} to the RB protocol, rendering them undetectable through standard ASF analysis.

\subsection{{Monotonicity of ASF and witness of quantum memory}}
A deviation in the ASF curve from a single-parameter exponential decay is often interpreted as a signature of non-Markovian dynamics \cite{PRXQuantum.2.040351,FigueroaRomero2022towardsgeneral,PhysRevA.89.062321,PhysRevA.93.022303,PhysRevA.92.022326,PhysRevA.103.022607,ceasura2022nonexponentialbehaviourlogicalrandomized,gandhari2025quantumnonmarkoviannoiserandomized,Mavadia2018,Veldhorst2014-db}. A definitive indicator is an \emph{increase} in the ASF with sequence length, since under the assumption Markovian noise, the ASF is $\bar{F}^{(m)}_M=A\prod_{t=1}^{m} q_t + B$ where the decay parameters $\{q_t\}$ correspond to CPTP map $\mathcal{N}_t$ and is determined by Eq.~\eqref{2nd mom coeff}, i.e.,
\begin{equation}
    q_t=c_{2}(\mathcal{N}_t) = \frac{\operatorname{Tr}\!\left(\kket{\mathbb{I}}\bbra{\mathbb{I}}\, \Proj{\mathcal{N}_t}\right) - 1}{d^2 - 1},
    \label{eq:mono_decay_para}
\end{equation}
 where we have used $\operatorname{Tr}\!\left(\Proj{\mathcal{N}_t}\right)=d$ since $\mathcal{N}_t$ is CPTP. Although the RB decay parameter $q_t$ can in principle take values in the interval $[-1/(d^2-1),1]$, a negative value requires the noise map to be sufficiently far from the identity channel. This notion of distance from identity channel can be quantified by the diamond distance.

 \begin{definition}
     \label{diamond distance}
     The diamond distance between two channels $\mathcal{M}$ and $\mathcal{N}$ is 

     \begin{equation}
         \label{dd_def}
         \Diamond\left(\mathcal{M},\mathcal{N}\right)=\frac{1}{2}\max_{\chi\in  \mathcal{H}_{\text{aux}}\otimes\mathcal{H}_S} \| \mathcal{I} \otimes \mathcal{N}(\chi)-\mathcal{I} \otimes \mathcal{M}(\chi)\|_{1}
     \end{equation}
    where $\chi$ is a quantum state living on the tensor product of system Hilbert space $\mathcal{H}_{S}$ and an auxiliary Hilbert space $\mathcal{H}_{aux}$. Since we work in finite-dimensional Hilbert spaces we can take $dim(\mathcal{H}_S)=dim(\mathcal{H}_{\text{aux}})$. The trace norm is defined as $\|A\|_1=\operatorname{Tr}\sqrt{A^{\dagger}A}$. The diamond distance gives a well defined notion of distinguishability between two channels in terms of single-shot discrimination tasks\cite{Sacchi_2005,Watrous_2018}.
     
 \end{definition}
 
 A sufficient condition for non-negativity of decay parameters is provided in the following theorem.
 \begin{theorem}
    \label{thm:diamond-sufficient-positive}
    For a Markovian error model, if the worst-case error, quantified by the diamond distance between the noise map $\mathcal{N}_t$ and identity channel, satisfies the condition,
    \begin{equation}
\Diamond(\mathcal I,\mathcal N_t)< \frac{d^2-1}{d^2}.
\label{eq:diamond-sufficient-positive}
\end{equation}
    then the corresponding decay parameters $q_t$ are guaranteed to be nonnegative.
\end{theorem}
\begin{proof}
    Consider the normalized Choi states  $\rho_{\mathcal I}=\kket{\mathbb I}\bbra{\mathbb I}/d,$ and $\rho_{\mathcal N_t}=\Proj{\mathcal N_t}/d,$
so that $\rho_{\mathcal I}$ is a pure state. The entanglement fidelity of the noise map $\mathcal{N}_t$ is given by \cite{NIELSEN2002249},
\begin{equation}
F_e(\mathcal N_t)
= \operatorname{Tr}(\rho_{\mathcal I}\rho_{\mathcal N_t})
= \frac{1}{d^2}\operatorname{Tr}\!\left(\kket{\mathbb I}\bbra{\mathbb I}\,\Proj{\mathcal N_t}\right),
\end{equation}
and the decay parameter in Eq. ~\eqref{eq:mono_decay_para} is obtained in terms of entanglement fidelity as
\begin{equation}
q_t=\frac{d^2 F_e(\mathcal N_t)-1}{d^2-1}.
\label{eq:qt-fe}
\end{equation}
Since $\rho_{\mathcal I}$ is pure, the Fuchs--van de Graaf inequality implies \cite{fuchs1998cryptographicdistinguishabilitymeasuresquantum}
\begin{equation}
1-F_e(\mathcal N_t)
\leq \frac{1}{2}\|\rho_{\mathcal I}-\rho_{\mathcal N_t}\|_1.
\end{equation}
Using the relation between the trace norm of Choi matrices and the diamond norm (See lemma 7 of \cite{Wallman_2014}),
\begin{equation}
\frac{1}{2}\|\rho_{\mathcal I}-\rho_{\mathcal N_t}\|_1
=
\frac{1}{2d}\left\|\kket{\mathbb I}\bbra{\mathbb I}-\Proj{\mathcal N_t}\right\|_1
\leq \Diamond(\mathcal I,\mathcal N_t),
\end{equation}
we obtain
\begin{equation}
F_e(\mathcal N_t)\geq 1-\Diamond(\mathcal I,\mathcal N_t).
\label{eq:fe-lower-bound}
\end{equation}
Using Eq.~\eqref{eq:fe-lower-bound} and  Eq.~\eqref{eq:qt-fe}, we obtain the the following lower bound
\begin{equation}
q_t \geq 1-\frac{d^2}{d^2-1}\,\Diamond(\mathcal I,\mathcal N_t),
\label{eq:qt-diamond-lower-bound}
\end{equation}
leading to the sufficiency condition in Eq. \eqref{eq:diamond-sufficient-positive} for non negativity of the decay parameters.
\end{proof}
In particular, for qubit systems this condition reduces to a threshold of 3/4, so in any well-calibrated experiment, where the diamond error is well below this value, the decay parameters are expected to be positive. Also note that the threshold in ~\eqref{eq:diamond-sufficient-positive} asymptotically approaches unity as system size increases.

For CCC-type errors, whose ASF is given by Eq.~\eqref{Eff.Mod.ASF.CCC}, we model the noise as a convex combination of Markovian error channels represented by CPTP maps $\mathcal{N}_x$. \textit{A priori}, there is no requirement that the decay parameter $q_x$ associated with a given Markovian branch be strictly positive. For instance, consider the action of a single Clifford gate $\mathcal{C}$ applied to a pure state $\ketbra{\psi}{\psi}$,
\begin{equation}
\mathcal{C}\!\left(\ketbra{\psi}{\psi}\right) 
= \mathcal{G} \circ 
\Bigg(\sum_{x} p_x \,\mathcal{N}_x\!\left(\ketbra{\psi}{\psi}\right)\Bigg),
\end{equation}
where $\mathcal{G}$ denotes the ideal gate implementation, $p_x$ are the branch probabilities, and $\mathcal{N}_x$ the associated error channels. The average fidelity of the error channel is given by $\sum_{x}p_x \int\bra{\psi}\mathcal{N}_x(\ketbra{\psi}{\psi})\ket{\psi} d\psi=\sum_xp_x\left(q_x+(1-q_x)/d\right)$. Since physical gates are expected to have high average fidelity, we impose the condition $\sum_xp_x\left(q_x+(1-q_x)/d\right)=1-\epsilon$, for $0<\epsilon\ll 1$, and therefore $\sum_x p_xq_x=(1-d\epsilon)/(d-1)$. Thus, even though the overall gate fidelity may be close to unity, the individual $q_x$ may in principle take negative values for some $x$. Such unfavorable branches will have correlations along the entire sequence albeit with small probability and may lead to oscillatory behavior in the ASF.

A toy example for such a case could be a convex combination of two Markovian branches $\mathcal{N}_1=\mathcal{I}$ with $q_1=1$ and $\mathcal{N}_2=\mathcal{X}$ with $q_2=-1/3$ where $\mathcal{X}$ is a bit-flip operation. If we target the average fidelity of single gate application to be $0.9$, the mixture weight on the bit-flip branch is $0.15$. The ASF for an $m$ length sequence is $A\left(0.85+0.15(-1/3)^{m+1}\right)+B$ which exhibits exponentially damped even–odd oscillations. Although this is a contrived example, it illustrates that the ASF for CCC noise models need not be monotonic even though single gate action is taken to be high fidelity. In practice however, such oscillatory deviations will be strongly suppressed by the exponential damping factor and may be further obscured by statistical fluctuations arising from finite sampling and the use of a limited Clifford gate subset. 

Moreover, for a CCC type model, if we make a reasonable physical assumption that for all noise maps $\mathcal{N}_x$, labeled by a Markovian branch $x$, the threshold in ~\eqref{eq:diamond-sufficient-positive} holds, i.e., each noise map is sufficiently close to identity channel, then decay paramaters $q_x$ are positive for all $x$. This is justified by observations; for instance, in \cite{PhysRevA.92.022326}, where the non-exponential decays observed in silicon quantum-dot qubit were explained by low-frequency drift of qubit resonance frequency which fits the CCC-type error model \cite{Veldhorst2014-db}. As long as the decay parameters lie in the range $(0,1)$, it follows that $\bar{F}^{(m)}_{CCC} >\bar{F}^{(m+1)}_{CCC}\;\;\forall m$ which is evident from
\begin{equation}
    \label{monotonic ASF CCC}
    \bar{F}^{(m)}_{CCC}-\bar{F}^{(m+1)}_{CCC}=A\left(\sum_{x}p_{x}q_{x}^{m+1}\left[1-q_x\right]\right)
\end{equation}
where parameters $\{p_x\} \in [0,1]$ are branch weights and $A=\operatorname{Tr}\!\left(\mathcal{M}(\ketbra{0}{0})\left[\mathcal{P}(\ketbra{0}{0})-\mathbb{I}/d\right]\right)$.
In typical RB settings $A>0$, reflecting that SPAM errors (captured by $\mathcal{P}$ and $\mathcal{M}$) are small perturbations around the ideal preparation and measurement, and an explicit sufficient condition for $A>0$ can be derived using the same type of threshold argument as above.

\begin{corollary}
\label{cor:Monotonicity_CCC}
For a CCC noise model, under the assumption that the noise map corresponding to every Markovian branch $x$ satisfies the sufficiency condition in Eq.~\eqref{eq:diamond-sufficient-positive}, the ASF, $\bar{F}^{(m)}_{CCC}$ is a monotonically decreasing function of the sequence length $m$.
\end{corollary}

Analogously, for CFF noise models, the monotonicity condition is 

\begin{gather}
\bar{F}^{(m)}_{CFF} - \bar{F}^{(m+1)}_{CFF}
= A \left( \sum_{\vec a,\vec x} \prod_{i=0}^{m} \gamma_{a_i,x_i \mid \vec{a}_{i-1}, \vec{x}_{i-1}} \right) \nonumber\\
\quad\times \left( 1 - \sum_{a_{m+1},x_{m+1}} \gamma_{a_{m+1},x_{m+1} \mid \vec{a}_{m},\vec{x}_{m}} \right),
\label{monotonic ASF CFF}
\end{gather} 
where $\gamma_{a_i,x_{i}|\vec{a}_{i-1},\vec{x}_{i-1}}=p(x_{i}|\vec{a}_{i-1},\vec{x}_{i-1}) \beta_{a_i|x_i}$. Since $p(x_{i}|\vec{a}_{i-1},\vec{x}_{i-1})$, the relevant decay parameters are $\beta_{a_i|x_i}$. These parameters can take values in the range $\left[-\operatorname{Tr}\left(\Proj{\mathcal{N}_{a_i|x_i}}\right)/d(d^2-1),1 \right]$ with the constraint $\sum_{a_i}\operatorname{Tr}\left(\Proj{\mathcal{N}_{a_i|x_i}}\right)=d$. Consequently, the ASF can, in principle, also exhibit non-monotonic behavior under CFF noise. However, as in the CCC case, such deviations are likely to be difficult to observe in experiments. 

\begin{corollary}
\label{cor:Monotonicity_CFF}
For a CFF noise model, assume that for every time step $t$ and every instrument setting/outcome pair $(x_t,a_t)$, the normalized CP map
$\mathcal{N}_{a_t|x_t}/kd$ (where $k=\operatorname{Tr}\left(\Proj{\mathcal{N}_{a_t|x_t}}\right)$)
is sufficiently close to the identity in the sense that
\begin{equation}
\Diamond\!\left(\mathcal I,\frac{\mathcal{N}_{a_t|x_t}}{kd}\right)
< \frac{d^2-1}{d^2}.
\label{eq:diamond_CFF-sufficient-positive}
\end{equation}
Then the ASF $\bar{F}^{(m)}_{\mathrm{CFF}}$ is a monotonically decreasing function of the sequence length $m$.
\end{corollary}

In \cite{Wallman_2014}, it was observed that any non-monotonic behaviour of the ASF constitutes a witness for non-Markovianity. The following corollary provide a refined version of the claim, namely a witness for the coherent memory in the environment.   

\begin{corollary}
\label{Cor:QM_witness}
Under the assumptions of Corollaries~\ref{cor:Monotonicity_CCC} and~\ref{cor:Monotonicity_CFF}, the ASF predicted by CCC and CFF (classical-memory) noise models is a monotonically decreasing function of the sequence length $m$. Consequently, an experimentally observed \emph{non-monotonic} ASF is incompatible with these classical-memory models in the above regime, and therefore points to either a violation of the stated assumptions or to additional non-Markovian effects due to quantum memory.
\end{corollary}

\subsection{Complete blindness to RB}\label{Complete blindness to RB}

In general, the ASF for classical memory noise models deviates from a single exponential decay, making it crucial to resolve and distinguish the different decay parameters present in the RB data. Even in cases where distinct decay parameters can be resolved (with methods like ESPRIT), non-Markovian noise may still mislead optimization strategies. Consider, for instance, a CCC-type model that is a convex combination of two Markovian branches with parameters $\{p,q_1\}$ and $\{1-p,q_2\}$. Suppose $q_1>q_2$ and one attempts to tune experimental parameters to bias the system toward the higher-fidelity branch (i.e., $p\approx1)$. While this tuning increases the average sequence fidelity, the physical nature of the corresponding error channels is essential, for instance, consider a scenario where $q_1$ corresponds to a coherent unitary rotation, whereas $q_2$  corresponds to a stochastic noise process. It is well known that coherent errors accumulate more adversely than stochastic errors with respect to worst-case error metrics \cite{Sanders_2016}. As a result, simply maximizing average fidelity may not yield the most robust error performance.

On the other hand consider the case where the decay parameters associated with the two Markovian branches are nearly identical, $q_1 \approx q_2=q$. In this situation, the ASF reduces to a form that appears indistinguishable from a purely Markovian process with a single decay parameter $q$. Nevertheless, the underlying noise remains non-Markovian and can lead to non-trivial effects, which we discuss in Sec.~\ref{Worst case errors}. This indicates that for non-Markovian errors with classical memory, if all decay parameters are equal, the ASF is \emph{completely blind} to the underlying non-Markovian correlations. 

As noted earlier, classical memory effects can arise from system-environment interactions where the environment itself is quantum mechanical. A natural question, therefore, is whether there exist class of system-environment interactions that generate classical memory noise, which is completely blind to the RB protocol. The following theorem provides a sufficient class RB blind Hamiltonian.

\begin{theorem}
\label{thm:RB_blind_Hamiltonians}
Consider a system--environment Hamiltonian of the form
\[
H^{ES}=\sum_i H_i^{E}\otimes H_i^{S}.
\]
Assume the environment operators commute, $[H_i^{E},H_j^{E}]=0$ for all $i,j$, with a common eigenbasis $\{\ket{\lambda}\}$ and eigenvalues $\{\lambda_i\}$. Then
\[
H^{ES}=\sum_{\lambda}\ketbra{\lambda}{\lambda}\otimes H_\lambda^{S},
\qquad
H_\lambda^{S}:=\sum_i \lambda_i H_i^{S},
\]
and the induced multi-time process is CCC (a convex mixture of Markovian branches labelled by $\lambda$; see Appendix~\ref{AppenB}). Moreover, if the RB decay parameters of all branches coincide, i.e.,
\[
q_\lambda = q_{\lambda'}\quad \text{for all }\lambda,\lambda',
\]
then the resulting CCC process is \emph{completely RB-blind}.
\end{theorem}

\begin{proof}
Since $[H_i^{E},H_j^{E}]=0$ for all $i,j$, the operators $\{H_i^{E}\}_i$ admit a common eigenbasis $\{\ket{\lambda}\}$ with eigenvalues $\{\lambda_i\}$. Expanding $H^{ES}$ in this basis gives
\[
H^{ES}
=\sum_{\lambda}\ketbra{\lambda}{\lambda}\otimes H_\lambda^{S},
\]
where $H_\lambda^{S}:=\sum_i \lambda_i H_i^{S}$. Tracing over the environment therefore yields a convex mixture of Markovian branches labelled by $\lambda$, i.e., a CCC process (see Appendix~\ref{AppenB}).

If the RB decay parameters of all branches coincide, $q_\lambda=q_{\lambda'}$ for all $\lambda,\lambda'$, then the ASF collapses to a single exponential with a unique decay parameter. Consequently, the RB signal is identical to that of a Markovian model with the same decay, and RB is completely blind to the underlying CCC correlations.

To make the equal-decay condition explicit, let $\{e_{\lambda,m}\}$ denote the spectrum of $H_\lambda^{S}$. For the unitary branch $U_\lambda(t)=e^{-itH_\lambda^{S}}$ (with $\hbar=1$), the corresponding RB decay parameter can be written as
\[
q_{\lambda}=\frac{1}{d^{2}-1}\left(\left|\operatorname{Tr}\left(U_\lambda(t)\right)\right|^{2}-1\right).
\]
Equivalently, $q_\lambda$ depends on spectral differences through quantities such as
\[
C_\lambda:=\sum_{m>n}\cos\!\bigl(t(e_{\lambda,m}-e_{\lambda,n})\bigr),
\]
so a sufficient condition for complete RB-blindness is that $C_\lambda$ is invariant over $\lambda$.
\end{proof}

 For instance, consider the case when the system and environment both are qubits and the coupling between them is given by $Z\otimes Z$. In this scenario, $H^{ES}=\ketbra{0}{0}\otimes Z - \ketbra{1}{1}\otimes Z $ and clearly the above condition is satisfied resulting in identical decay constants for both the Markovian branches in the  CCC process. Clearly, such Hamiltonians lead to non-Markovian noise which is invisible in RB protocol. As we will show in the following section, the memory effects in such processes still have a non-trivial impact on the worst-case error, an important parameter to assess fault tolerance.  

\section{Worst case errors}\label{Worst case errors}
RB protocols are designed to estimate the average error rates of gates, and the protocols are experimentally accessible. However, fault-tolerance thresholds are typically determined by worst-case error metrics, such as the diamond norm, which are not directly measurable in standard experiments \cite{PhysRevLett.117.170502,PhysRevA.85.042311,Wallman_2014}. In particular, for coherent errors, worst-case and average metrics can differ by orders of magnitude \cite{Sanders_2016}. It is therefore essential to analyze worst-case errors in the presence of non-Markovian noise, especially in scenarios where the non-Markovianity is completely hidden from RB, as shown in the previous section.

Consider a qubit system-environment interaction between successive gate applications described by $U^{ES}=\operatorname{exp}\left(-i \delta\; Z \otimes Z\right)$ and $\delta=\pi/100$. Let the initial environment state be 
\begin{equation}\label{eq:initstate}
\ket{\phi}_{E}=\sqrt{p}\ket{0}+\sqrt{1-p}\ket{1}.
\end{equation}
This interaction produces two Markovian branches with identical decay parameters $q_1=q_2=q=\left(|\operatorname{Tr}\left(\operatorname{exp}(-i\delta Z)\right)|^2-1\right)/3=1-4\delta^{2}/3 +\mathcal{O}(\delta^4)$. Consequently, the ASF exhibits a single exponential decay and the average gate error inferred from such a model is $r_{\text{avg}}\approx2\delta^2/3 +\mathcal{O}(\delta^4)$.

\begin{figure}
    
    \centering
    \includegraphics[width=1.0\columnwidth]{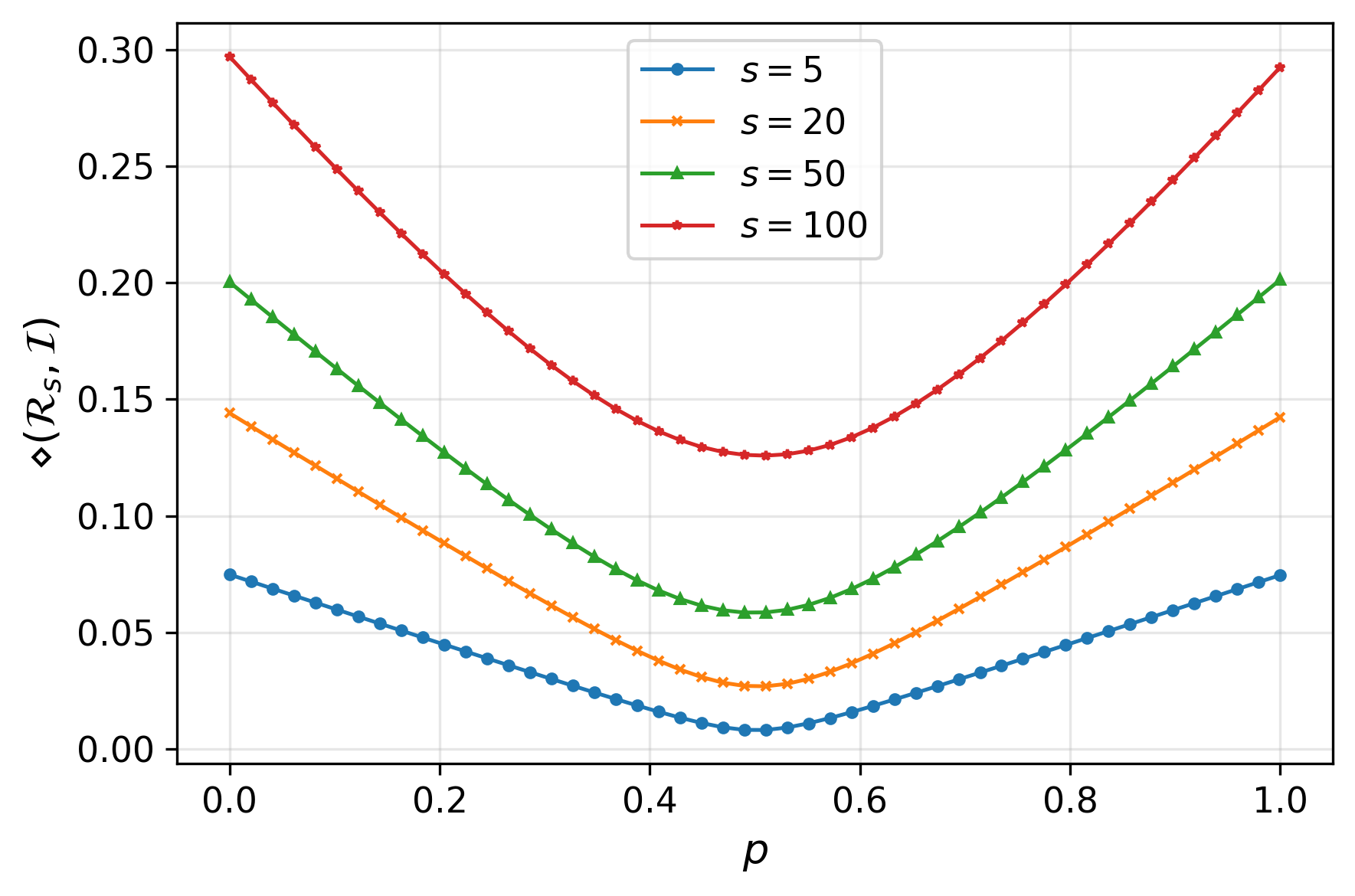}
    \caption{Diamond norm distance average for various sequence lengths with respect to mixing parameter $p$ in \eqref{eq:initstate}. }
\label{fig:diamond_dis}
\end{figure}

Given a sequence of gates in the RB procedure $S_{\alpha} \equiv \{\mathcal{G}^{\alpha}_{1},\mathcal{G}^{\alpha}_{2} \cdots,\mathcal{G}^{\alpha}_{m},\mathcal{G}^{\alpha}_{m+1}\}$ and re-defining the Clifford operations $\mathcal{K}_{1}^{\alpha}=\mathcal{G}_{1}^{\alpha}$ and $\mathcal{K}_{i}^{\alpha}=\mathcal{G}_{i}^{\alpha} \circ \mathcal{K}_{i-1}^{\alpha}~ ~\forall~ i \geq 2$ as in Sec.~\ref{ASF calculation using process matrix formalism}, the entire sequence of gates is a CPTP map $\mathcal{R}_{S_{\alpha}}(\rho)=\sum_{x \in \{1,2\}} p_{x} \mathcal{K}^{\alpha\dagger}_{m}\circ \mathcal{N}_{x} \circ \mathcal{K}^{\alpha}_{m} \cdots\circ\mathcal{K}^{\alpha \dagger}_{1}\circ \mathcal{N}_{x}\circ \mathcal{K}^{\alpha}_{1}\circ \mathcal{N}_{x}(\rho)$ such that 
\begin{equation}\label{noise map comp}
\begin{aligned}
\mathcal{N}_{1}(\rho) &= \operatorname{exp}\left(-i \delta Z\right) \rho \operatorname{exp}\left(i \delta Z\right)   \\
\mathcal{N}_{2}(\rho) &= \operatorname{exp}\left(i \delta Z\right) \rho \operatorname{exp}\left(-i \delta Z\right)
\end{aligned}
\end{equation}
and $p_1=p$, $p_2=1-p$. To quantify the worst-case error, we consider the diamond distance between $\mathcal{R}_{S_{\alpha}}$ and the identity channel $\mathcal{I}$ which represents ideal gate implementations
\begin{equation}
    \label{diamond distance}
    \Diamond\left(\mathcal{R}_{S_{\alpha}},\mathcal{I}\right)=\frac{1}{2}\max_{\chi\in  \mathcal{H}_{\text{aux}}\otimes\mathcal{H}_S} \| \mathcal{I} \otimes \mathcal{R}_{S_{\alpha}}(\chi)-\mathcal{I} \otimes \mathcal{I}(\chi)\|_{1}
\end{equation}
The diamond distance in Eq.~\eqref{diamond distance} also depends explicitly on the chosen gate sequence $S_{\alpha}$. In the presence of temporal correlations, the noise cannot be assigned independently to individual gates, and thus error metrics such as the diamond distance can only be meaningfully defined at the level of an entire sequence rather than per gate. Let $\Lambda_{\chi}^{\alpha}=\mathcal{R}_{S_{\alpha}}\otimes \mathcal{I}(\chi)-\mathcal{I} \otimes \mathcal{I}(\chi)$ and up-to second order in $\delta$ we have $\operatorname{exp}\left(-i \delta Z\right)=\mathbb{I}(1-\delta^2/2)-iZ\delta+\mathcal{O}(\delta^3)$. Using this we get
\begin{equation}
    \label{Lamda}
    \Lambda^{\alpha}_{\chi}=i(2p-1)\delta \Lambda^{\alpha}_{1,\chi} - \delta^{2} \Lambda^{\alpha}_{2,\chi}+\mathcal{O}(\delta^3)
\end{equation}
such that 
\begin{equation}
\begin{aligned}
    \label{Lambda1}
    \Lambda^{\alpha}_{1,\chi}&=\sum_{j=1}^{m} \left[\chi,\mathbb{I}\otimes P_{j}^{\alpha}\right]+\left[\chi,\mathbb{I}\otimes Z\right]\\
    \Lambda^{\alpha}_{2,\chi}&=(m+1) \chi +\sum_{j=1}^{m} \left(\mathbb{I}\otimes P_{j}^{\alpha}Z\right)\chi + \sum_{j=1}^{m} \chi\left(\mathbb{I}\otimes ZP_{j}^{\alpha}\right) +\\
    &\sum_{j>k} \left(\left( \mathbb{I}\otimes P_{j}^{\alpha}P_{k}^{\alpha}\right)\chi+\chi\left(\mathbb{I}\otimes P_{j}^{\alpha}P_{k}^{\alpha}\right)\right) -\\
    &\left(\sum_{j=1}^{m}\mathbb{I} \otimes P_{j}^{\alpha}+\mathbb{I}\otimes Z\right) \chi \left(\sum_{k=1}^{m}\mathbb{I} \otimes P_{k}^{\alpha}+\mathbb{I}\otimes Z\right)
\end{aligned}
\end{equation}
where $P_{j}^{\alpha}=K^{\alpha \dagger}_{j} Z K^{\alpha }_{j}$ and $[A,B]$ is the commutator operation. The diamond distance in Eq.~\eqref{diamond distance} now becomes $\max_{\chi\in \mathcal{H}_S \otimes \mathcal{H}_{S}}\|\sqrt{\Lambda_{\chi}^{\alpha\dagger}\Lambda^{\alpha}_{\chi}}\|_{1}$. When $p=1$ or $p=0$ (purely Markovian coherent errors), $\Lambda^{\alpha}_{\chi}=\pm i\delta\Lambda^{\alpha}_{1,\chi}+\mathcal{O}(\delta^2)$ whereas for $p=0.5$ (maximal mixing between the two Markovian branches), $\Lambda^{\alpha}_{\chi}=-\delta^{2}\Lambda^{\alpha}_{2,\chi}$. Although calculating Eq.~\eqref{diamond distance} exactly requires knowledge of the specific sequence $S_{\alpha}$, these results suggest scalings:
\begin{equation}
    \label{scaling}
\begin{aligned}
    p=\{0,1\} \rightarrow&\Diamond\left(\mathcal{R}_{S_{\alpha}},\mathcal{I}\right) \approx \mathcal{O}(\delta)=\mathcal{O}(\sqrt{r_{\text{avg}}})\\
    p=0.5\rightarrow&\Diamond\left(\mathcal{R}_{S_{\alpha}},\mathcal{I}\right) \approx \mathcal{O}(\delta^2)=\mathcal{O}(r_{\text{avg}})   
\end{aligned}
\end{equation}
We verified this numerically. In order to investigate the variation of worst case error for different  sequence lengths with respect to the mixing parameter $p$, we exploit the fact that the diamond norm distance between two channels can be cast in the form of a semidefinite program \cite{10.1088/978-0-7503-3343-6}. In particular, for a given sequence $S_{\alpha}$ and mixing parameter $p$, the diamond norm distance of the resulting effective channel $\mathcal{R}_{S_{\alpha}}$ from the identity channel is obtained as 
\begin{align}
\text{max}\quad 
  & \tfrac12\,\operatorname{Tr}\!\Big[X\big(\Proj{\mathcal{R}_{S_\alpha}}-\Proj{\mathcal{I}}\big)\Big] \\
\text{s. t.}\quad 
  & -2(\mathbb{I}\otimes \rho) \le X \le 2(\mathbb{I}\otimes \rho),\nonumber \\
  & \rho \succeq 0, \quad \operatorname{Tr}[\rho]=1 .\nonumber
\end{align}
For a given sequence length, we sample 50 random Clifford sequences and find the average diamond norm distance corresponding to mixing parameter $p$. We plot the results for various sequence lengths for mixing parameter ranging from zero to one in Fig.~\ref{fig:diamond_dis}. 

As evident from the graph, the diamond norm distance is minimum for the parameter value $p=0.5$, i.e., for the maximal CCC process, and maximum for $p=0,1$, i.e., when the process is a single Markovian branch with coherent errors. Importantly, this result suggests that, for present transmon-based quantum hardwares, where a natural $ZZ$ coupling is present, it is useful to initiate the environmental qubits in the superposed state of ground and exited state (such as $|+\rangle$) rather than the ground state for reduced worst-case errors. More broadly, our findings indicate that non-Markovian effects are not necessarily detrimental to gate performance. Understanding how non-Markovian memory effects influence the worst-case error remains an interesting direction for further investigation.

\section{Discussion and Conclusion}
\label{Discussion and Conclusion}

In this work, we developed a general framework to study randomized benchmarking (RB) under non-Markovian noise models that introduce classical temporal correlations, either through stochastic classical fields or interactions with a quantum environment. We derived analytical expressions for the average sequence fidelity (ASF) in such settings and demonstrated that, unlike the standard Markovian case, the ASF can be expressed as a sum of multiple exponential decays. We showed how the additional decay parameters can be interpreted to yield useful information for estimating average errors, and how established signal-processing techniques such as ESPRIT and MUSIC  can be employed to extract these parameters from RB data.

Our analysis revealed several important features. First, while ASF curves with classical correlations can in principle exhibit non-monotonic behavior, such deviations are exponentially suppressed, hence difficult to observe in practice. When all decay parameters remain positive, the ASF is guaranteed to decrease monotonically, making non-Markovian effects invisible to monotonicity-based diagnostics. Thus, any observed violation of a monotonically decreasing ASF is a strong indicator for genuinely quantum memory effects. Moreover, we identified classes of interaction Hamiltonians that generate classical memory processes completely blind to RB, yielding ASF curves indistinguishable from Markovian noise despite the presence of temporal correlations. Finally, we investigated the implications for worst-case error metrics. While RB provides estimates of average gate errors, fault-tolerance thresholds are governed by worst-case performance quantified by the diamond norm \cite{PhysRevLett.117.170502,Sanders_2016,PhysRevA.89.062321}. We argued that classical memory processes can alter worst-case error behavior, and in some scenarios non-Markovian effects may even reduce worst-case errors. These findings underline the importance of going beyond average metrics when assessing the reliability of quantum devices especially in the presence of correlated noise.

Overall, our results clarify both the capabilities and blind spots of randomized benchmarking in the presence of temporal correlations. They also motivate the development of complementary protocols capable of jointly estimating average and worst-case error metrics under realistic noise models \cite{zhang2025learning,Sanders_2016,Wallman_2014}. Moreover, it is important to come up with theoretical bounds on worst-case errors in case of non-Markovian noise, similar to the bounds presented in \cite{Wallman_2014} for Markovian noise models. It would also be interesting to explore whether the RB protocol can be modified to reliably provide an operational metric on the amount of non-Markovian noise present during gate implementations. Finally, we note that the present analysis has been limited to Clifford operations under the assumption of gate-independent errors. We anticipate that our framework can be readily extended to more general gate sets, and that established techniques \cite{FigueroaRomero2022towardsgeneral,Wallman2018randomized} can be naturally integrated to treat these broader scenarios.

\section*{Acknowledgment}
V.S, A.K.R, S.M and J.K acknowledge funding from the Sydney Quantum Academy. This work was supported by the Hon-Hai Research Institute through the Australian Quantum Software Network (AQSN) micro-grant. We also wish to acknowledge support from the ARC Centre of Excellence for
Engineered Quantum Systems, EQUS (CE170100009), and SK acknowledges support through a EQUS Deborah Jin
Fellowship.

\appendix
\section{Link product and constructing time-ordered process matrix}\label{AppendA}

 Here, we provide a brief summary of the link product and the use of it to construct multi-time processes.\\
The link product ($\star$) of two operators $A \in \mathcal{L}\left( \bigotimes_{p_j \in p} \mathcal{H}_{p_j} \right)$ and \( B \in \mathcal{L}\left( \bigotimes_{q_j \in q} \mathcal{H}_{q_j} \right) \) is defined as 
\begin{align}
A \star B = \operatorname{Tr}_{p \cap q} \left[ \left( \mathbb{I}_{q \setminus p} \otimes A^{T_{p \cap q}} \right) \left( B \otimes \mathbb{I}_{p \setminus q} \right) \right] \notag,
\\
\in \mathcal{L} \left( \mathcal{H}_{q \setminus p} \otimes \mathcal{H}_{p \setminus q} \right),
\end{align}
where the operators $(.)^T_{p\cap q}$  and $\operatorname{Tr}_{p\cap q}$ represent the partial transpose and partial trace, respectively, over the common space $p\cap q$. In particular, when the operators do not share common space, link product reduces to the usual tensor product, i.e., $A\star B = A\otimes B$ when $p\cap q = \phi$. The link product has the following useful properties. It is (i) Hermiticity preserving, i.e., $A,B$ Hermitian $\implies A\star B$ is Hermitian, (ii) positivity preserving, i.e., $A,B\geq 0\implies A\star B\geq 0$, (iii) commutative $A\star B = B\star A$ and (iv) associative, i.e., $A\star (B\star C) = (A\star B)\star C)$, where operators $A,B,$ and $C$ do not have any joint overlap.\\ Note that the action of a map $\mathcal{M}:S_I\rightarrow S_O$ on state $\rho^{S_I}$ in terms of the link product is
\begin{equation}\label{link_channel}
\begin{aligned}
    \mathcal{M}(\rho^{S_I}) &= \operatorname{Tr}_{S_I}(\Proj{\mathcal{M}}^{S_IS_O}(\rho^{T_{S_I}}\otimes \mathbb{I}^{S_O})\\
    &=\Proj{\mathcal{M}}^{S_IS_O}\star\rho^{S_I} \\&= \rho^{S_I}\star\Proj{\mathcal{M}}^{S_IS_O}
\end{aligned}
\end{equation}

Consider a multi-time scenario with $N$ interventions depicted by the labs $A_1, A_2,\cdots,A_N$. In general, the initial system and environment can be correlated and given by the joint state $\rho^{A_1^{I}E_1}$, where we use the superscript $I (O)$ to denote the input (output) space of the labs. The local operations in the lab $A_i$ are elements of the instrument $\{\mathcal{M}_{a_i}:A_i^I\rightarrow A_i^O\}$, and between the labs $A_i$ and $A_{i+1}$, there is a joint system environment interaction given by the unitary $\mathcal{U}_i:E_iA_i^O\rightarrow E_{i+1}A^I_{i+1}$. With this setting, the probability of obtaining the outcomes $a_1,a_2,\cdots a_N$ is given by the Born rule,
\begin{equation}
\begin{aligned}
     \mathbb{P}(a_1,a_2,\cdots,a_N)= \operatorname{Tr}_{E_NA_N^O}&\big[\mathcal{M}_{a_N}(\mathcal{U}_{N-1}\cdots\\&\cdots(\mathcal{U}_2(\mathcal{M}_{a_2}(\mathcal{U}_{1}(\mathcal{M}_{a_1}(\rho))))].
\end{aligned}
\end{equation}
Using Eq.~\eqref{link_channel} recursively, and using the commutativity and associativity properties of the link product, one can rearrange the above to obtain the following.
\begin{equation}\label{Born_link}
\begin{aligned}
     \mathbb{P}(a_1,a_2,\cdots,a_N) = \operatorname{Tr}[W^T\cdot M],
\end{aligned}
\end{equation}
where $W$ and $M$ are operators in the space $\bigotimes_i(A_i^I\otimes A_i^O)$ of the following form
\begin{equation}
    \begin{aligned}
        W = \rho^{E_1A_1^I}\star\big(\bigstar_{i=1}^{N-1}\Proj{\mathcal{U}}^{E_iA_i^OE_{i+1}A_{i+1}^I}\big)\star\mathbb{I}^{E_N},
    \end{aligned}
\end{equation}
and
\begin{equation}
    M = \bigotimes_i \Proj{\mathcal{M}_{a_i}}^{A_i^IA_i^O}.
\end{equation}
The object $W$ is known as the process matrix and is a bounded positive operator. This mapping of multi-time processes in terms of positive operator allows one to employ tools from many body systems for investigating multi-time processes. For example, in the Markovian scenario, where the system is initialized in state $\rho^{A_1^I}$, followed by the application of channels $\mathcal{E}_1,\cdots, \mathcal{E}_{n-1}$ is given by
\begin{equation}
    W = \rho_0 \ \star \Proj{\mathcal{E}_1} \ \star\cdots \star\Proj{\mathcal{E}_{n-1}} \star \mathbb{I}.
\end{equation}
Since there are no shared spaces, the link product reduces to the usual tensor product, resulting in the process matrix of the following form.
\begin{equation}
    W = \rho_0 \ \otimes \Proj{\mathcal{E}_1}\otimes \cdots \otimes \Proj{\mathcal{E}_{n-1}} \otimes \mathbb{I}.
\end{equation}
That is, the process matrix of a Markovian process is similar to an uncorrelated state. A process which cannot be factorized as above is called a non-Markovian process.

\section{Hamiltonian description of non-Markovian processes with classical memory}\label{AppenB}

As mentioned in the main text, the classical memory process can result from interaction of the system with a quantum environment. For instance, consider the following class of environment-system Hamiltonian
\begin{equation}
    H^{ES} = \sum_{x}H_{x}^{E}\otimes H^{S}_{x},
\end{equation}
such that $[H_{x}^{E},H_{y}^{E}] = 0 \;\;\forall x, y$ \cite{Goswami_2025}. The constraint on the environment part of the Hamiltonian implies that there is a common eigenbasis $\{|\lambda\rangle\}$ for the set $\{H^{E}_x\}$, so one can rearrange $H^{ES}$ to the form $H^{ES} = \sum_{\lambda} |\lambda\rangle\langle\lambda|^{E}\otimes H^{S}_{\lambda}$, where $H_{\lambda}^{S} = \sum_{x}\lambda^{x}H_{x}^{S}$ and we will replace the Hilbert space notation $\mathcal{H}^{S/E}$ with simply $S/E$ for convenience. If the interaction time is $\Delta t$, the resulting unitary in the Choi form is given by $\Proj{U}^{ESE'S'} = \sum_{\lambda,\lambda'}|\lambda\lambda\rangle\langle\lambda'\lambda'|^{EE'}\otimes |\tilde{U}_{\lambda}\rangle\langle\tilde{U}_{\lambda'}|^{SS'}$, where $\tilde{U}_{\lambda} = \exp{(-i\Delta t H_\lambda})$.  Now consider a three intervention scenario with labs $A, B$ and $C$, and the interactions between the labs active for $t_{AB}$ and $t_{BC}$, respectively and let $A_{I/O}$ denote the input/output Hilbert space for lab $A$. If the process starts with an arbitrary correlated environment-system state $\rho^{E_{1}A_{I}}$, the resulting process matrix is given by,
\begin{equation}
\begin{aligned}
    W = \rho^{E_1A_I}\star\Proj{&U({t_{AB})}}^{E_1A_OE_2B_I}\\&\star \Proj{U(t_{BC})}^{E_2 B_OE_3C_I}\star\mathbb{I}^{E_3}
    \end{aligned}.
\end{equation}
where $A_i,B_i,C_i$ denote the corresponding system Hilbert space. Using the Choi form of the interaction unitaries result in the following,
\begin{equation}
    \begin{aligned}
        W = \sum_{\lambda\lambda'\sigma\sigma'}&(\rho^{E_1A_I}\star|\lambda\lambda\rangle \langle \lambda'\lambda'|^{E_1E_2}\star|\sigma\sigma\rangle\langle\sigma'\sigma'|^{E_2E_3}\\&\star\mathbb{I}^{E_3})\otimes|\tilde{U}_{\lambda}\rangle\langle\tilde{U}_{\lambda'}|^{A_OB_I}\otimes|\tilde{U}_{\sigma}\rangle\langle\tilde{U}_{\sigma'}|^{B_OC_I},
    \end{aligned}
\end{equation}
where we have used the fact that the link product between objects without common spaces is the usual tensor product. A straightforward simplification results in the following form,
\begin{equation}
    \begin{aligned}
        W &= \sum_{\lambda}(\rho^{A_IE_1}\star|\lambda\rangle\langle\lambda|^{E_1})\otimes\Proj{\tilde{U}_{\lambda}}^{A_OB_I}\otimes\Proj{\tilde{U}_{\lambda}}^{B_OC_I}\\
        &=\sum p_{\lambda }W_{\lambda},
    \end{aligned}
\end{equation}
where $W_{\lambda}$ is Markovian process with the weight $p_\lambda = Tr(\rho^{A_IE_1}\star|\lambda\rangle\langle\lambda|^{E_1})$. Therefore, the above class of Hamiltonian results in a convex sum of Markovian processes, i.e., CCC process regardless of the amount of initial correlation in the environment-system state. The proof is straightforward to extend to arbitrary number of time-steps and can be generalized to include time-dependence in the Hamiltonian acting between the labs, for more details, we refer the readers to \cite{Goswami_2025}. Importantly, this class of Hamiltonian includes the generator of the CNOT gate, and Ising interactions such as $ZZ$ coupling, which is an always-on interaction between superconducting qubits in present quantum hardwares \cite{PRL2022_ZZcoupling,Mundada2019_ZZ_Coupling}. 

In a similar manner, one can obtain a classical feed-forward memory process through interaction with quantum environments. This case however, is not as straightforward as the CCC case, where the interaction was with a common environment at every time step. For instance, again consider a three intervention scenario, with the interaction during the first time step of the form $H_1 = \sum_m H^{E_0}_m \otimes H^{E_1S}\otimes \mathbb{I}^{E_2}$ and the interaction during the second time step of the form $H_2 = \sum_m H_m^{E_0}\otimes H_{m}^{E_1}\otimes H_{m}^{E_2S}$. If the following conditions are satisfied, 
\begin{enumerate}
    \item $[H^{E_0}_{m},H^{E_0}_{n}]=0$ for all $m,n$ and for both $H_1$ and $H_2$
    \item $[H^{E_1}_{m},H^{E_1}_{n}]=0$ for all $m,n$ for $H_2$
    \item Initial system-environment state is uncorrelated in spaces $E_1$ and $E_2$
\end{enumerate}

then it is straightforward to show that the resulting process matrix is
of the form,
\begin{equation}
    \begin{aligned}
        W = \sum_{xy} p_x\rho_x^{A_I}\otimes\Proj{\mathcal{E}_{y|x}}^{A_OB_I}\otimes\Proj{\mathcal{T}_{xy}}^{B_OC_I}.
    \end{aligned}
\end{equation}
Here, $\Proj{\mathcal{E}_{y|x}}$ is CJ map  corresponding to the CP map $\mathcal{E}_{y|x}$ with $\sum_y\mathcal{E}_{y|x} $ being a CPTP map for all $x$, and $\Proj{\mathcal{T}_{xy}}$ are CJ maps corresponding to CPTP maps, and $p(x)\rho_x^{A_I}$ has its origin in the reduced states of the initial $ES$ state in some basis $|x\rangle$ of Hilbert space of $E_0$. Evidently, the process is not a classical common cause process but a more general classical memory process with feed-forward.

\section{{ASF calculation in process matrix framework}} \label{addn.appen.ASF}

\subsection{Derivation of the ASF}

Here we provide a step-by-step derivation of Eq.~\eqref{avg.seq.fidelity_short}. Consider a RB protocol with $m+1$ gates where the first $m$ gates are picked randomly from the Clifford group of $n$ qubits, $\mathcal{CF}_n$ and the last gate is the motion reversal gate. For a given sequence of gates $S_{\alpha}$ each noisy gate $\mathcal{C}^{(\alpha)}_t$, applied at time-step $t$, is decomposed into the noisy system-environment unitary and the ideal gate as $\mathcal{C}_{t}\equiv \left(\mathcal{I} \otimes \mathcal{G}_{t}\right)\circ\mathcal{U}^{SE}_{t} $. The process matrix is then evaluated as 

\begin{equation}
    \label{appen_eq: process matrix}
    W = (\rho \otimes \sigma) \star \Proj{\mathcal{U}^{SE}_1} \star \cdots \star \Proj{\mathcal{U}^{SE}_{m+1}} \star \mathbb{I}^{E},
\end{equation}
where the identity at the end of the equation is equivalent to tracing out the environmental degree of freedoms. 

For this derivation it would be helpful to keep an account of the relevant Hilbert spaces.  Let $(\rho \otimes \sigma) \in \mathcal{L}(\mathcal{H}_{0}^{S_O}\otimes \mathcal{H}^{E})$ and $\Proj{\mathcal{U}^{SE}_t} \in \mathcal{L}(\mathcal{H}_{t-1}^{S_O} \otimes \mathcal{H}^{E} \otimes \mathcal{H}_{t}^{S_I} \otimes \mathcal{H}^{E})$, where the label $S_{O},S_{I}$ refer to the output/input Hilbert spaces of the system of interest at time-step $t$. Note that although we have not labeled the environment with temporal labels, they should also be contracted in the correct order. Now notice that in the RB protocol, the first unitary doesn't get twirled therefore we can define a new system-environment state $\lambda = (\rho \otimes \sigma) \star \Proj{\mathcal{U}^{SE}_1}$. This can lead to SPAM errors influencing the temporal correlations. The process matrix therefore is a positive linear operator $W \in \mathcal{L}(\mathcal{H}_{1}^{S_I} \otimes \mathcal{H}_{1}^{S_O} \otimes \cdots\otimes \mathcal{H}_{m}^{S_I} \otimes \mathcal{H}_{m}^{S_O} \otimes \mathcal{H}_{m+1}^{S_I})$. 

Similarly, the instruments in the form of ideal Clifford gates can be composed into a higher order operator as 

\begin{equation}
    \label{appen_eq:instrument}
    \mathbb{T}_\alpha = \Proj{\mathcal{G}^{(\alpha)}_1} \otimes \cdots \otimes \Proj{\mathcal{G}^{(\alpha)}_m} \otimes \Proj{\mathcal{G}^{(\alpha)}_{m+1}} \star E^T 
\end{equation}

where $\mathbb{T}_\alpha \in \mathcal{L}(\mathcal{H}_{1}^{S_I} \otimes \mathcal{H}_{1}^{S_O} \otimes \cdots\otimes \mathcal{H}_{m}^{S_I} \otimes \mathcal{H}_{m}^{S_O} \otimes \mathcal{H}_{m+1}^{S_I})$ is another positive linear operator that encodes the operations that were intended to be applied in the noiseless limit. Note that the link product is a tensor product if the operators don't share common Hilbert spaces.

Each ideal gate can be represented in the Choi form $\Proj{\mathcal{G}^{(\alpha)}_t} \in \mathcal{L}(\mathcal{H}_{t}^{S_I} \otimes \mathcal{H}_{t}^{S_O})$ and the Choi representation of the POVM element is $E^T \in \mathcal{L}(\mathcal{H}_{m+1}^{S_O})$ can be contracted with the last motion reversal gate as $\Proj{\mathcal{G}^{(\alpha)}_{m+1}} \star E^T=\left(G^{\dagger(\alpha)}_{m+1}EG^{(\alpha)}_{m+1}\right)^T$. We then re-define the Clifford gates as $\mathcal{K}_{1}^{(\alpha)}=\mathcal{G}_{1}^{(\alpha)}$ and $\mathcal{K}_{i}^{(\alpha)}=\mathcal{G}_{i}^{(\alpha)} \circ \mathcal{K}_{i-1}^{(\alpha)}~ ~\forall~ i \geq 2$. Then for any time $t\geq 2$ we can write 

\begin{equation}
\label{appen_eq:gate redefinition}
\begin{aligned}
\Proj{\mathcal{G}^{(\alpha)}_t}
&= \Proj{\mathcal{K}_{t}^{(\alpha)} \circ \mathcal{K}_{t-1}^{\dagger(\alpha)}} \\
&= \kket{K_{t}^{(\alpha)}K_{t-1}^{\dagger(\alpha)}} \bbra{K_{t}^{(\alpha)}K_{t-1}^{\dagger(\alpha)}} \\
&= K_{t-1}^{*(\alpha)} \otimes K_{t}^{(\alpha)} \,
   \kket{\mathbb{I}}\bbra{\mathbb{I}} \,
   K_{t-1}^{T(\alpha)} \otimes K_{t}^{\dagger(\alpha)}
\end{aligned}
\end{equation}

where for an operator $A$, its complex conjugate is $A^*$. We can re-write Eq.~\eqref{appen_eq:instrument} can be written as 

$$
\mathbb{T}_\alpha = \mathbb{K}^T \left(\kket{\mathbb{I}}\bbra{\mathbb{I}}\otimes E^T\right) \mathbb{K}^{*} 
$$

where 

$$
\mathbb{K}^{(\alpha)}
:=\mathbb{I}~\!\bigotimes_{t=1}^{m}\!\left({K}_{t}^{T(\alpha)} \otimes {K}_{t}^{\dagger(\alpha)} \right)
$$

The sequence fidelity of a given sequence $S^{\alpha}$ is then the contraction of the process matrix with the instrument operator

\begin{equation}
\label{appen_eq: Seq.Fid}
\begin{aligned}
F_{\alpha}^{(m)}
&= W * \mathbb{T}_{\alpha} \\
&= \operatorname{Tr}\left(
    W^T \mathbb{K}^T
    \left(\kket{\mathbb{I}}\bbra{\mathbb{I}} \otimes E^T\right)
    \mathbb{K}^{*}
\right) \\
&= \tilde{W}_{\alpha} \star \Gamma
\end{aligned}
\end{equation}

where $\tilde{W}_{\alpha}=\mathbb{K}^{(\alpha)}\,W\,\mathbb{K}^{\dagger(\alpha)}$ and $\Gamma=\kket{\mathbb{I}}\bbra{\mathbb{I}}^{\otimes (m)} \otimes E^T$.

The average sequence fidelity is the average of sequence fidelity $F_{\alpha}^{(m)}$ over uniformly selected gates from the Clifford group. This expression is given in Eq.~\eqref{avg.seq.fidelity_short}

\begin{equation}
    \label{appen_eq:avg.seq.fid}
    \bar{F}^{(m)}=\frac{1}{\Omega^m}\sum_{\alpha=1}^{\Omega^m}\tilde{W}_{\alpha} \star {\Gamma},
    \quad \Omega=|\mathcal{CF}_n|.
\end{equation}

\subsection{Unitary 2-designs and Haar second moment}

Consider an operator $O \in \mathcal{L}(\mathcal{H} \otimes \mathcal{H})$ then the Haar second moment is defined as 

\begin{equation}
\label{appen_eq:2ndmom}
\begin{aligned}
\mathbb{E}_{U \sim \mu_H}\!\left[ U^{\otimes 2} O (U^\dagger)^{\otimes 2} \right]
&= \int_{U} d\mu_H \left[ U^{\otimes 2} O (U^\dagger)^{\otimes 2} \right] \\
&= c_{\mathbb{I},O}\,\mathbb{I} + c_{\mathbb{F},O}\,\mathbb{F}.
\end{aligned}
\end{equation}
where
\begin{equation}
\label{appen_eq:coeff}
\begin{aligned}
c_{\mathbb{I},O}
&= \frac{\operatorname{Tr}(O) - d^{-1}\operatorname{Tr}(\mathbb{F}O)}{d^2 - 1}, \\
c_{\mathbb{F},O}
&= \frac{\operatorname{Tr}(\mathbb{F}O) - d^{-1}\operatorname{Tr}(O)}{d^2 - 1}.
\end{aligned}
\end{equation}

where $\mathbb{F}$ is the SWAP operator and $d$ is the dimension of Hilbert space $\mathcal{H}$. 

This is a consequence of Schur-Weyl duality which relates the second moment over a Haar measure to the corresponding permutation matrices acting on $\mathcal{H} \otimes \mathcal{H}$, namely $\{\mathbb{I},\mathbb{F}\}$. The Clifford group is particularly useful because it forms a unitary 2-design, which means that its second moment under the uniform distribution matches the Haar second moment \cite{Mele2024introductiontohaar}.

In Eq.~\eqref{modified_mark_ASF} we saw that the ASF for the Time-dependent Markovian model is

\begin{equation}
\label{appen_eq:1stASF}
\bar{F}^{(m)}_M=\left(\tilde{\rho}\otimes Z_2 \otimes \cdots \otimes Z_{m+1}\right)\star \Gamma,
\end{equation}

where 

$$
Z_t:=\frac{1}{\Omega}\sum_{K_t\in\mathcal{CF}_n}\bigl(K_{t}^{T(\alpha)} \otimes K_{t}^{\dagger(\alpha)}\bigr)\,\Proj{\mathcal{N}_t}\,\bigl(K_{t}^{T(\alpha)} \otimes K_{t}^{\dagger(\alpha)}\bigr)^{\dagger}
$$
and $\Omega$ is the number of elements in the Clifford group of $n$ qubits. Now $Z_t \in \mathcal{L}(\mathcal{H}_{t-1}^{S_O} \otimes \mathcal{H}_{t}^{S_I})$ then taking a partial transpose over the Hilbert space $\mathcal{H}_{t-1}^{S_O}$ gives

\begin{equation}
    \label{appen_eq:Z_t}
    Z_t^{T_{S_O}}:=\frac{1}{\Omega}\sum_{K_t\in\mathcal{CF}_n}\bigl(K_{t}^{\dagger(\alpha)} \otimes K_{t}^{\dagger(\alpha)}\bigr)\,\Proj{\mathcal{N}_t}\,\bigl(K_{t}^{(\alpha)} \otimes K_{t}^{(\alpha)}\bigr)
\end{equation}

which is equal to the Haar second moment. Since partial transpose of $\mathbb{F}=\kket{\mathbb{I}}\bbra{\mathbb{I}}$, it is straightforward to verify that

\begin{equation}
    \label{appen_eq:Z_t new}
    Z_t= c_{1}(\mathcal{N}_t)\,\frac{\mathbb{I}}{d} \;+\; c_{2}(\mathcal{N}_t)\,\kket{\mathbb{I}}\bbra{\mathbb{I}}
\end{equation}

where the coefficients 

\begin{equation}
    \label{appen_eq:new_coeff}
    \begin{aligned}
c_{1}(\mathcal{N}_t) &= \frac{d\,\operatorname{Tr}\!\left(\Proj{\mathcal{N}_t}\right) -  \operatorname{Tr}\!\left(\kket{\mathbb{I}}\bbra{\mathbb{I}}\, \Proj{\mathcal{N}_t}\right)}{d^2 - 1}, \\
c_{2}(\mathcal{N}_t) &= \frac{\operatorname{Tr}\!\left(\kket{\mathbb{I}}\bbra{\mathbb{I}}\, \Proj{\mathcal{N}_t}\right) - d^{-1}\operatorname{Tr}\!\left(\Proj{\mathcal{N}_t}\right)}{d^2 - 1}.
\end{aligned}
\end{equation}

are related as $c_{1}(\mathcal{N}_t)+c_{2}(\mathcal{N}_t)=\operatorname{Tr}\!\left(\Proj{\mathcal{N}_t}\right)/d$. For any CPTP map $\operatorname{Tr}\!\left(\Proj{\mathcal{N}_t}\right)=d$ and therefore $c_{1}(\mathcal{N}_t)+c_{2}(\mathcal{N}_t)=1$. Eq.~\eqref{appen_eq:1stASF} can then be evaluated to give the ASF 

$$
\bar{F}^{(m)}_M=A\prod_{t=1}^{m} c_{2}(\mathcal{N}_t) + B
$$
where $A=\operatorname{Tr}\!\left[E\left(\mathcal{N}_{1}(\rho)-\mathbb{I}/d\right)\right]$ and $B=\frac{1}{d}\operatorname{Tr}(E)$.
 
\section{Data fitting for Randomised Benchmarking}
\label{AppenC}
\begin{figure}[h!]
    \centering
    \includegraphics[width=1.05\linewidth]{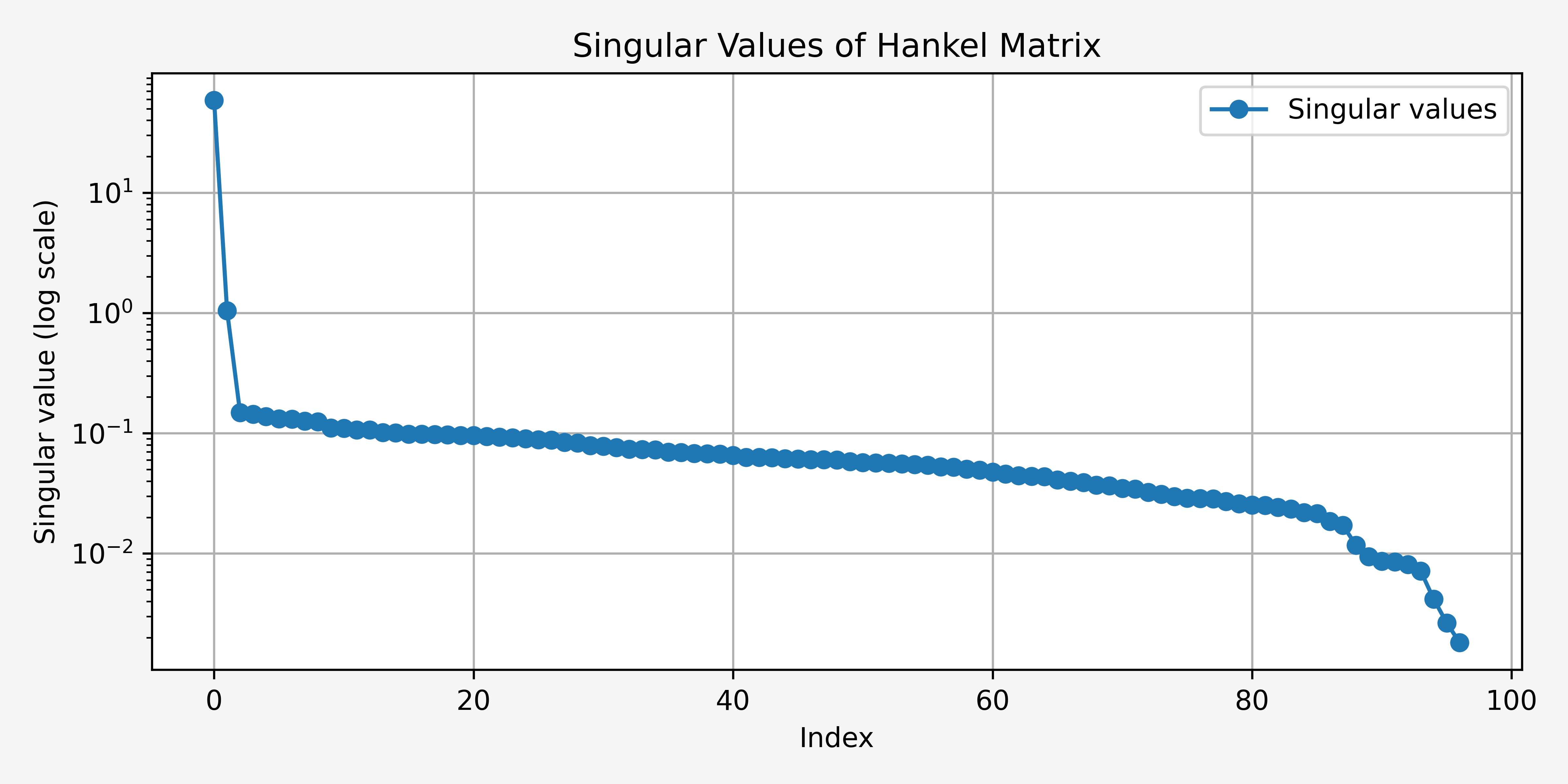}
    \caption{The plot shows singular values of the Hankel matrix from the simulated data presented in Sec.~\ref{Complete blindness to RB}. The singular values are plotted in logarithmic scale in the Y-axis, and the two dominant singular values are clearly distinguished from the rest which correspond to two different expopnents present in the data.}
    \label{fig:Singular values}
\end{figure}

Here, we present our data fitting algorithm and examples of extracting the decay parameters from RB data for different CCC processes, where the origin of the RB data is a model comprising of two exponential decay terms and additional noise: 
\begin{equation}
    y_i = A_1q_1^{i+1}+A_2q_2^{i+1}+B+n_i,
\end{equation}
where $A_1=Ap_1$ and $A_2=Ap_2$. The constant terms $A$ and $B$ capture the information of SPAM errors, while $p_1$ and $p_2$
denote the probabilities corresponding to the two different Markovian branches of the CCC process; and $n_i$ is a noise term representing the deviation from the true ASF due to finite sampling or other possible noise. The fitting program aims to extract the probabilities $p_1$, $p_2$, the constants $A, B$, and the exponents $q_1$, $q_2$ from the provided data \{$y_i$\}. For data fitting, we employ ESPRIT technique followed by solving a linear equation to extract the required parameters.
The complete workflow comprises of the following steps:
\begin{enumerate}
    \item The ASF data for a sequence length $m$ $[y_1, y_2, \cdots, y_m]$ can be expressed in the following matrix equation:
\begin{equation*}
    \begin{pmatrix}
       y_1\\
       y_2\\
       \vdots\\
       y_m
    \end{pmatrix}  
    = \begin{pmatrix}
       q_1^2 & q_2^2 & 1\\
       q_1^3 & q_2^3 & 1\\
       \vdots & \vdots &\vdots\\
       q_1^{m+1} & q_2^{m+1} & 1
    \end{pmatrix}
    \begin{pmatrix}
        A_1\\
        A_2\\
        B
    \end{pmatrix}
    +\begin{pmatrix}
        n_1\\
        n_2\\
        \vdots\\
        n_m
    \end{pmatrix}
\end{equation*}

    \item Splitting and writing the data in the form of a Hankel matrix of dimension $l \times (m-l+1)$:  
\begin{multline}\label{Hankel equation}
    \begin{pmatrix}
        y_1 & \cdots & y_{m-l+1}\\
        \vdots & \ddots & \vdots\\
        y_l & \cdots & y_m
    \end{pmatrix}
    = \begin{pmatrix}
        q_1^2 & q_2^2 & 1\\
        \vdots & \vdots & \vdots\\
        q_1^{l+1} & q^{l+1}_2 & 1
    \end{pmatrix}
    \begin{pmatrix}
        A_1 &\cdots &A_1q_1^{m-l}\\
        A_2 &\cdots &A_2q_2^{m-l}\\
        B &\cdots& B
    \end{pmatrix}\\
    +\begin{pmatrix}
      n_1 & \cdots & n_{m-l+1}\\
      \vdots  & \ddots & \vdots\\
      n_l & \cdots & n_m
    \end{pmatrix},
\end{multline}
we obtain its singular values. Eq.~(\ref{Hankel equation}) can be rewritten in shorthand as $Y=VA+N$, where $Y$ is the Hankel matrix, $V$ is called the Vandermonde matrix and $N$ is the noise matrix. We choose $l=\lfloor m/2\rfloor$, where $\lfloor\cdot\rfloor$ is the floor function.
The dominant singular values of the Hankel matrix denotes the number of different exponents (or decay parameters) present in the data (see Fig. (\ref{fig:Singular values})).
\item The number of dominant singular values (supplied as the ``model order'') as well as the data serve as input for the ESPRIT algorithm, which outputs the different exponents.
\item Using the exponents, we determine $A_1$, $A_2$ and constant $B$ by solving the linear equation Eq.~\eqref{Hankel equation}. 
\item Once we obtain the parameter values for $A_1 \text{ and } A_2$, we can find $p_1 \text{ and } p_2$ from the following relations: $p_1=\frac{A_1}{A_1+A_2}$ and $p_2=\frac{A_2}{A_1+A_2}$.
\end{enumerate}
ESPRIT algorithm is based on a rotational invariance property of the Vandermonde matrix. This relation takes the form 
\begin{equation}\label{Rotational Invariance}
    J_2V=J_1 V H
\end{equation}
where $J_1$ and $J_2$ are the last and first row deletion matrix respectively, and $H$ is the diagonal matrix comprising the exponents.
\begin{align*}
    J_1 &=\scriptsize{\begin{pmatrix}
    1 & 0 & \cdots &0 &0\\
    0 & 1 & \cdots &0 &0\\
    \vdots & \vdots & \ddots &0 &0\\
    0 & 0 & \cdots &1 &0
\end{pmatrix}_{l-1\times l}} = [\mathbf{I}_{l-1}~~\mathbf{0}]\\
J_2&=\scriptsize{\begin{pmatrix}
    0 & 1 & 0&\cdots  &0\\
    0 & 0 & 1&\cdots  &0\\
    \vdots & \vdots & \vdots &\ddots  &0\\
    0 & 0 & 0&\cdots  &1
\end{pmatrix}_{l-1\times l}} = [\mathbf{0} ~~ \mathbf{I}_{l-1}]\\
H &= \scriptsize{\begin{pmatrix}
    q_1 & 0 & 0\\
    0 & q_2 & 0\\
    0 & 0 & 1
\end{pmatrix}}
\end{align*}
The singular value decomposition of the Hankel matrix in Eq.~(\ref{Hankel equation}) can be broken into two parts and selecting the dominant singular values as the signal and the rest as noise, we obtain $Y=U_s\Sigma_sW_s^\dagger+U_n\Sigma_nW_n^\dagger$. Here, $\Sigma_s$ contain the dominant singular values that are identified as signal and $U_s, W_s$ are the corresponding part of the isometries $U, W$ which constitute the singular value decomposition. Similarly, $U_n,W_n$ correspond to the noise part. Comparing it to $Y=VA+N$, one can write the signal part as $VA=U_s\Sigma_sW_s^\dagger$ or $V=U_s\Sigma_sW_s^\dagger A^{-1}$. Replacing this relation in Eq.~(\ref{Rotational Invariance}), it follows that
\begin{equation}\label{ESPRIT}
    J_2U_s=J_1U_s\Phi
\end{equation} where the eigenvalues of $\Phi$ corresponds to the exponents of the data. The ESPRIT algorithm uses Eq.~(\ref{ESPRIT}) and upon solving the linear equation for an unknown $\Phi$, gives out the eigenvalues (exponents $q_1, q_2$). For the ideal case with no noise and sequence length 200, it can perfectly resolve different exponents as with difference as close as $10^{-4}$. In presence of noise the performance depends both on the noise strength as well as the separation between the exponents (see \cite{Helsen_2022_General} for details).

\bibliography{apssamp}% Produces the bibliography via BibTeX.

\end{document}